\documentclass[reprint,amsmath,amssymb,superscriptaddress,nofootinbib,aps]{revtex4-2}

\usepackage{xcolor}

\usepackage{graphicx}
\usepackage{braket}
\usepackage{caption}
\usepackage{subcaption}
\usepackage{array}
\usepackage{xr}
\usepackage{tikz}
\usetikzlibrary{positioning, arrows.meta}
\usepackage{amsthm}
\usepackage{mathtools}
\usepackage{comment}
\usepackage{hyperref}

\usepackage[ruled,linesnumbered]{algorithm2e}
\SetKwInput{KwIn}{Input}
\SetKwInput{KwOut}{Output}
\SetKwInput{KwRes}{Resources}
\DontPrintSemicolon

\newtheorem{Def}{Definition}
\newtheorem{Prop}{Proposition}
\newtheorem{The}{Theorem}

\newtheorem{Cor}{Corollary}
\newtheorem{Lem}{Lemma}

\theoremstyle{definition}

\newcommand{\firstreview}[1]{\textcolor{black}{#1}}
\newcommand{\secondreview}[1]{\textcolor{black}{#1}}
\newcommand{\thirdreview}[1]{\textcolor{black}{#1}}

\begin{document}

\title{Quantum Circuits for the Metropolis-Hastings Algorithm}

\author{Baptiste Claudon}\email{baptiste.claudon@qubit-pharmaceuticals.com}
\affiliation{Qubit Pharmaceuticals, Advanced Research Department, 75014 Paris, France}
\affiliation{Sorbonne Universit\'e, LJLL, UMR 7198 CNRS, 75005 Paris, France}
\affiliation{Sorbonne Universit\'e, LCT, UMR 7616 CNRS, 75005 Paris, France}
\author{Pablo Rodenas-Ruiz}\email{pablo.rodenasruiz@epfl.ch}
\affiliation{Qubit Pharmaceuticals, Advanced Research Department, 75014 Paris, France}
\affiliation{École Polytechnique Fédérale de Lausanne (EPFL), 1015 Lausanne, Switzerland}
\author{Jean-Philip Piquemal}\email{jean-philip.piquemal@sorbonne-universite.fr}
\affiliation{Qubit Pharmaceuticals, Advanced Research Department, 75014 Paris, France}%
\affiliation{Sorbonne Universit\'e, LCT, UMR 7616 CNRS, 75005 Paris, France}%
\author{Pierre Monmarché}\email{pierre.monmarche@sorbonne-universite.fr}
\affiliation{Sorbonne Universit\'e, LJLL, UMR 7198 CNRS, 75005 Paris, France}%
\affiliation{Sorbonne Universit\'e, LCT, UMR 7616 CNRS, 75005 Paris, France}%
\affiliation{Institut Universitaire de France, 75005 Paris, France}

\begin{abstract}
Szegedy's quantization of a reversible Markov chain provides a quantum walk whose spectral gap is quadratically larger than that of the classical walk. Quantum computers are therefore expected to provide a speedup of Metropolis-Hastings (MH) simulations. Existing generic methods to implement the quantum walk require coherently computing the transition probabilities of the underlying Markov kernel. However, reversible computing methods require a number of qubits that scales with the complexity of the computation. This overhead is undesirable in near-term fault-tolerant quantum computing, where few logical qubits are available. In this work, we present a \secondreview{Szegedy} quantum walk construction which follows the classical proposal-acceptance logic, and does not require further reversible computing methods. \secondreview{We also compare this construction with an alternative to Szegedy's approach which also provides a quadratic gap amplification.} Since each step of the quantum walks uses a constant number of proposal and acceptance steps, we expect the end-to-end quadratic speedup to hold for MH Markov Chain Monte-Carlo simulations.
\end{abstract}

\maketitle

\section{Introduction}

Markov Chain Monte Carlo (MCMC) methods constitute a set of algorithms based on Markov processes to draw samples from a probability distribution. They have found application in diverse areas, from molecular dynamics~\cite{https://doi.org/10.1155/2015/183918}, with implications in drug design, to the training of neural networks in machine learning~\cite{doi:10.1126/science.220.4598.671}. These algorithms also serve as a tool of choice in the study of lattice and spin systems in statistical physics~\cite{10.5555/1051461} or the estimation of the volume of convex bodies~\cite{LOVASZ2006392}. The most widely used MCMC method is the Metropolis-Hastings algorithm~\cite{10.1093/biomet/57.1.97}. In order to sample from a distribution $\pi$ that is known only up to a multiplicative constant, the algorithm consists in simulating a Markov process with stationary distribution $\pi$. The Markov process is simulated through a succession of move proposals that are accepted with certain probabilities. These acceptance probabilities are chosen so that the Markov process is reversible with respect to the target $\pi$. The algorithm's cost is described by the number of simulation steps required to approach the stationary distribution, known as the mixing time. When the process is reversible, the mixing time is of the order of the inverse spectral gap of the chain (i.e. the difference between the modulus of its two largest eigenvalues).

Recent efforts have focused on designing algorithms based on the principles of quantum mechanics. In 2004, Szegedy showed that every reversible Markov chain can be mapped to a quantum walk with quadratically larger spectral gap~\cite{1366222}. In addition, the stationary distribution is invariant under the quantum walk dynamics. Combined with algorithms to reflect through eigenspaces~\cite{Kitaev1995QuantumMA, claudon2025simplealgorithmreflecteigenspaces} and amplitude amplification~\cite{Brassard_2002, Yoder_2014}, Szegedy's walk provides samples from the stationary distribution with a quadratically improved dependence on the spectral gap, compared to the classical mixing time.

Despite this reduction in the number of steps, the overall complexity of the walk is given by the number of steps times the cost of each step, which could make the quantum walk prohibitively expensive in practice. Generic implementation methods rely on the coherent computation of the transition probabilities of the underlying kernel (see~\cite{chiang2009efficientcircuitsquantumwalks} for an overview of existing methods and an efficient implementation in the sparse case). In the presence of additional symmetries, more efficient methods have been developed. For instance, there exist methods without coherent arithmetic for certain simple random walks on \thirdreview{certain graphs or specific symmetries}~\cite{PhysRevA.86.042338, Loke_2017}. Nonetheless, the MH algorithm does not require the computation of the transition probabilities, nor presents a group structure. Computing the diagonal elements of the Markov kernel, the probability of rejecting the proposed move can involve evaluating a sum of many terms. Each step of the quantum walk may be much more costly than a classical step, which could negate the quantum speedup.

The quantum walk of the MH kernel should therefore not be implemented by computing the transition probabilities. Instead, it should rely on steps of the proposal kernel and the computation of the move acceptance probability matrix. As explained in Appendix A of~\cite{lemieux2020efficient}, the difficult part of implementing this quantum walk lies in the rejection of certain samples. Indeed, the sample-forgetting step is not invertible, and is thus nontrivial to simulate using unitary dynamics. \thirdreview{Ozols et al. provide a quantum rejection sampling algorithm without using Markov chains~\cite{Ozols_2012}.} Lemieux et al. address the problem by following a different quantization procedure from Szegedy's~\cite{lemieux2020efficient}. The authors provide complete quantum circuits specifically designed to study the Ising model. Childs et al.~\cite{childs2022quantum} address the problem with reversible computing methods, closely following the classical move acceptance procedure~\cite{childs2022quantum}. Still, the qubit requirements of their technique are not fixed and depend on the acceptance probability matrix. Thus, the method might be discarded in the near term. 
\secondreview{In this work, we construct the quantum step operator of a Markov chain \thirdreview{defined} on pairs of states, for which one of the marginals follows exactly the MH transition probabilities. These quantum circuits allow us to avoid the coherent computation of the rejection probabilities, requiring only a constant number of calls to $O_T$ and $O_A$, the proposal and acceptance oracles. We also describe a method similar to that of~\cite{lemieux2020efficient} which does not rely on the MH quantum step operator and can be implemented with a constant number of calls to the same oracles (Appendix~\ref{app:other}). Our constructions minimize the number of qubits on which reversible arithmetic is performed. We compare them to the problem-dependent ancilla counts of previous methods in Table~\ref{tab:costs}.}

\thirdreview{The MH Markov chain evolves in two operations: proposal of a candidate state, and acceptance or rejection. Since rejection discards the proposal, this operation is not directly invertible. We address this by extending the state space with a single-state memory register. In case of acceptance, the memory stores the previous state. In case of rejection, the memory contains the rejected proposal. The proposal step itself remains independent of the memory register. This change of perspective makes the logic reversible, leading to a unitary quantum encoding of the process.} 

We show how to recover samples of $\pi$ from the stationary measure of this \thirdreview{extended} kernel. We also prove that the spectral gap of the dual kernel is as large as that of the original MH kernel. As a consequence, the quadratic speedup \thirdreview{of the corresponding quantum walk} is maintained. To the best of our knowledge, our construction is the only Szegedy quantization of MH kernels that does not rely on the coherent computation of their matrix elements or transition probabilities. It therefore suggests that MH quantum walks can always be implemented without prohibitive algorithmic overhead.

\section{Contributions}
\thirdreview{At the heart of our construction is the definition of a dual Markov kernel on the extended state space $\mathbb{S}^2$, which is introduced in Section~\ref{sec:duals}. The extended state space allows to record the information needed to make accept/reject updates compatible with a unitary implementation. More explicitly, from an edge $(x, y)\in \mathbb S^2$, the chain first proposes a new candidate $z\in \mathbb S$ as in the standard MH algorithm, reaching the intermediate state $(x, z)$. We then accept the state $z$ with the same acceptance probabilities as in the MH algorithm. In case of acceptance, the dual chain step leads to the edge $(z, x)$. In case of rejection, the new state is $(x, z)$. Thus, the second component of the pair stores either the previous state or the rejected proposal, while the first component is a realization of the MH chain, so its marginal distribution converges to $\pi$.}

\thirdreview{
In Section~\ref{sec:qstepops} we construct a low-resource quantum step operator of the dual kernel from projected unitary encodings (PUEs) of the proposal and acceptance steps. We prove that the dual kernel previously described may be written as a product $\mathcal {P=TA}$, where $\mathcal T$ and $\mathcal A$ are both reversible with respect to a common measure $\nu$ whose marginal on the first component is $\pi$. Since $\mathcal T$ and $\mathcal A$ do not commute, $\mathcal P$ itself is not reversible; we therefore also encode its time reversal $\mathcal P^\star=\mathcal{AT}$ and combine the two into a symmetric encoding to which the qubitized-walk spectral theorem applies.}

\thirdreview{Finally, Section~\ref{sec:gap} shows that this symmetric encoding has a spectral gap in $\Omega(\delta)$, where $\delta$ is the spectral gap of the original MH chain, so the resulting qubitized walk has angular gap in $\Omega\left(\sqrt{\delta}\right)$, preserving the quadratic amplification. Figure~\ref{fig:flowchart} summarizes the construction, while Algorithm~\ref{alg:sample_pi} specifies the sampling procedure and its resource requirements. The significantly reduced qubit requirements of our method allow for exact numerical simulations, which we use to illustrate the performance of our construction for the Metropolis-Adjusted Langevin Algorithm, of wide interest in molecular dynamics and machine learning.}

\begin{figure}[t!]
\centering

\colorlet{boxBorder}{black!35}
\colorlet{boxFill}{black!2!white}
\colorlet{inputBorder}{blue}
\colorlet{inputFill}{blue!6!white}
\colorlet{outputBorder}{red}
\colorlet{outputFill}{red!6!white}
\colorlet{arrow}{black!60}
\colorlet{mutedText}{black!55}

\providecommand{\fcap}[1]{{\itshape\footnotesize\color{mutedText} #1}}
\providecommand{\ftitle}[1]{{\bfseries\small #1}}

\begin{tikzpicture}[
  font=\small,
  every node/.style={text=black, inner sep=0pt},
  card/.style={
    rectangle,
    draw=boxBorder,
    fill=boxFill,
    line width=0.5pt,
    rounded corners=3pt,
    inner xsep=5.5pt,
    inner ysep=4pt,
    text width=64mm,
    align=center
  },
  cin/.style={
    card,
    draw=inputBorder,
    fill=inputFill,
    line width=0.6pt
  },
  cout/.style={
    card,
    draw=outputBorder,
    fill=outputFill,
    line width=0.6pt
  },
  arr/.style={
    -{Latex[length=2.5mm,width=1.6mm]},
    line width=0.7pt,
    draw=arrow
  }
]

\node[cin] (s1) {%
  \ftitle{Input}\\[2pt]
  proposal $T$, acceptance $A$, target $\pi$ on $\mathbb{S}$\\[1pt]
  quantum oracles $O_T,\,O_A$
};

\node[card, below=4mm of s1] (s2) {%
  \ftitle{Dual edge kernel}\\[2pt]
  edges $\mathcal{S} = \{(x,y) : T(x,y) > 0\}\subset\mathbb{S}^2$\\[2pt]
  $\mathcal{P} = \mathcal{T}\mathcal{A}$,\quad
  $\mathcal{P}^{\star} = \mathcal{A}\mathcal{T}$\\[2pt]
  $\nu(x,y) = \pi(x)\,T(x,y)$\\[2pt]
  \fcap{$\mathcal{T},\mathcal{A}$ reversible w.r.t.~$\nu$; Sec.~\ref{sec:duals}}
};

\node[card, below=4mm of s2] (s3) {%
  \ftitle{Step encodings}\\[2pt]
  PUEs $O,O_{\star}$ encode $\square,\square_{\star}$\\[1pt]
  $\square\ket{x,y} = \ket{x,y,\mathcal{P}((x,y),\cdot)}$\\[2pt]
  \fcap{built from $O_T$ and $O_{\mathcal A}$ derived from $O_A$; Lem.~\ref{lem:OmathcalA}, Props.~\ref{prop:construct_dual},~\ref{prop:construct_dual_star}}
};

\node[card, below=4mm of s3] (s4) {%
  \ftitle{Hermitianization}\\[2pt]
  $\overline{\mathcal P}
  = \ket{0}\!\bra{1}\!\otimes\!\mathcal P
  + \ket{1}\!\bra{0}\!\otimes\!\mathcal P^{\star}$\\[1pt]
  $\boxtimes
  = \ket{0}\!\bra{0}\!\otimes\!\square_{\star}
  + \ket{1}\!\bra{1}\!\otimes\!\square$\\[2pt]
  $\mathcal U = X\otimes S$\\[2pt]
  \fcap{$(\mathcal{U},\boxtimes)$ is a SPUE of the discriminant $\overline{\mathcal{D}}$ of $\overline{\mathcal{P}}$; spectral gap $\Omega(\delta)$; Prop.~\ref{prop:overline_dual_P}}
};

\node[card, below=4mm of s4] (s5) {%
  \ftitle{Qubitized quantum walk}\\[2pt]
  $\mathcal W=(2\boxtimes\boxtimes^{\dagger}-I)\,\mathcal U$\\[1pt]
  $1$-eigenstate in $\mathrm{range}(\boxtimes)$:
  $\boxtimes\ket{+,\nu,0}$\\[2pt]
  \fcap{angular gap $\Omega(\sqrt{\delta})$; Thm.~\ref{the:summary}}
};

\node[cout, below=4mm of s5] (s6) {%
  \ftitle{Output}\\[2pt]
map $\boxtimes\ket{+,\nu,0}\to\ket{+,\pi,0,0}$\\[1pt]
  measure the tail register\\[2pt]
  \fcap{via $O_T^\dagger O^\dagger
  \bigl(\ket{0}\!\bra{0}\otimes S + \ket{1}\!\bra{1}\otimes I\bigr)$;
  Prop.~\ref{prop:pi_and_plusnu}}
};

\draw[arr] (s1) -- (s2);
\draw[arr] (s2) -- (s3);
\draw[arr] (s3) -- (s4);
\draw[arr] (s4) -- (s5);
\draw[arr] (s5) -- (s6);

\end{tikzpicture}
\caption{
\thirdreview{Roadmap of the construction. The original Metropolis--Hastings chain is lifted to an edge-space chain, implemented quantumly by the $O_T, O_A$ oracles, Hermitianized, and then qubitized to obtain the final walk operator $\mathcal W$. }
}
\label{fig:flowchart}
\end{figure}

\section{Use case: Metropolis-Adjusted Langevin Algorithm}\label{sec:use_case}

The Metropolis-Hastings algorithm aims at obtaining samples from a probability distribution $\pi$. It proceeds in a succession of proposal and acceptance steps. When targeting a multi-dimensional continuous measure $\pi$, proposals are often obtained using a Langevin dynamics. Informally, the process is a gradient descent with random fluctuations. Let us define the process more precisely. Given an initial position $X_0=x_0$, the position at step $k+1$ is chosen using the position at step $k$ according to:
\begin{equation}
X_{k+1}=X_k+\tau\nabla\log(\pi(X_k))+\sqrt{2\tau}\xi_k,
\end{equation}
where $\tau>0$ is a small time step and the $(\xi_k)_k$ are independent and identically distributed standard Gaussian variables. The process resulting from adding the acceptance step is called Metropolis-Adjusted Langevin Algorithm (MALA).

In this work, we construct a quantum circuit acting on $4$ state registers and using $3$ ancillae to discretize the dynamics. Given a state space of size $n$, we only require $m = \lceil\log n \rceil$ qubits to label all the states. Therefore, our circuit uses a total of $4m+3$ qubits. To address the general context of molecular simulation, let us consider molecular dynamics of $N$-atom systems. The relevant state space is $\mathbb R^{3N}$, where each atom is described with three spatial coordinates. Thus, the qubit requirements of $12N$ coordinate registers and $3$ ancillae scale linearly with the number of atoms under study, offering a scalable path towards realistic systems. In addition to its low qubit requirements, our circuit does not make use of expensive quantum arithmetic operations. It only uses a constant number of quantum proposal and acceptance steps.

\begin{figure}
    \centering
    \includegraphics[width=\linewidth]{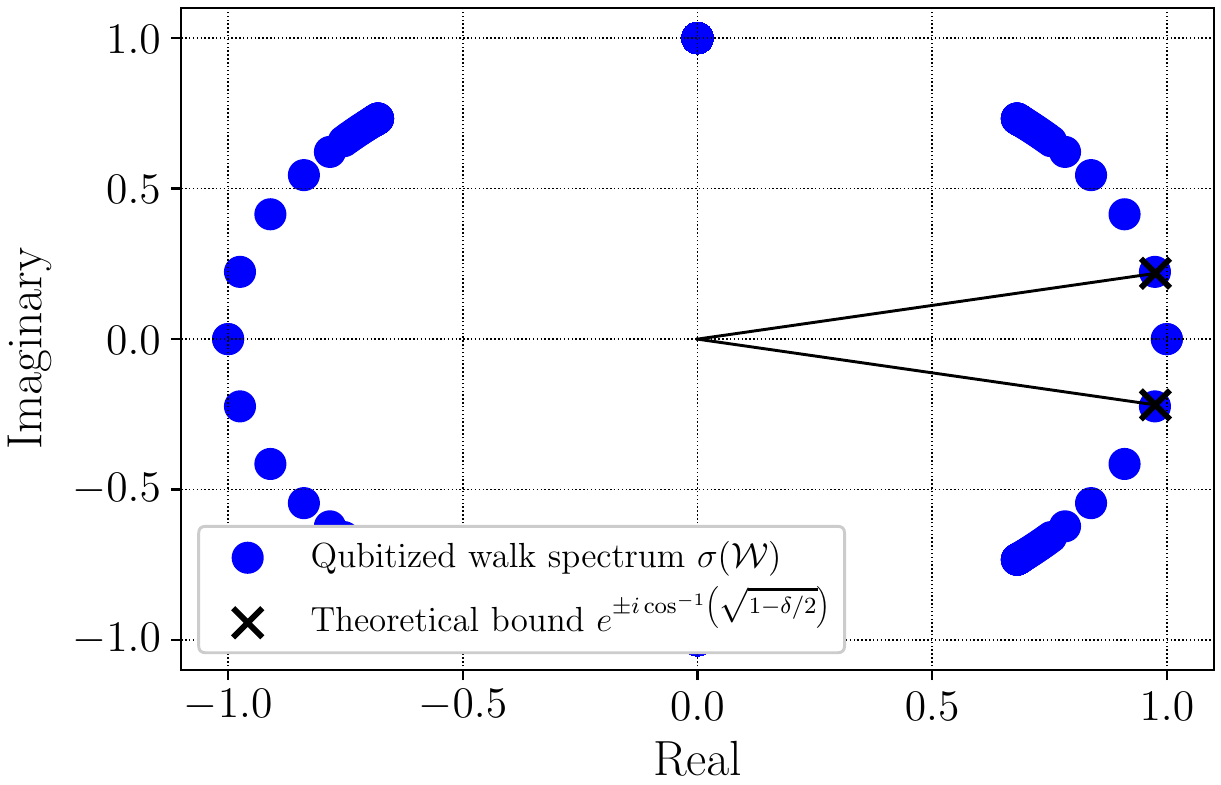}
    \caption{Spectral properties of the implemented walk operator $\mathcal W$. The process under study is the MALA\thirdreview{, with Metropolis choice,} targeting the measure $\pi\propto \exp(-U)$ for a two-well potential \thirdreview{$U:[-1, 1]\to\mathbb R$, $x\mapsto (x^4-x^2)/32$. The interval $[-1, 1]$ is discretized into $64$ equally spaced states}. $\delta$ is the spectral gap of the classical walk. $\mathcal W$ presents a quadratically amplified spectral gap.}
    \label{fig:example}
\end{figure}

For concreteness, consider a well-known toy model where $U$ is a one-dimensional two-well potential and $\pi$ is the corresponding Boltzmann distribution $\pi\propto e^{-U}$. In Appendix~\ref{app:OAOT}, we explain how to efficiently construct $O_T$ without ancilla and $O_A$ with a single ancilla. As shown in Figure~\ref{fig:example} for a total of 27 qubits (with state registers of size $m=6$, discretizing the dynamics with 64 classical states), the underlying unitary has a spectral gap that is quadratically larger than that of the classical process $\delta$. Indeed, its spectral gap is larger than the theoretical minimum gap of $\cos^{-1}\left(\sqrt{1-\delta/2}\right)\in \Omega\left(\sqrt\delta\right)$. This work therefore suggests the oracular quadratic speedup to hold for the Metropolis-Hastings algorithm in practical applications. It is supplemented by an open-source code~\cite{Rodenas_Ruiz_Claudon}. The code contains additional examples and allows to apply our construction to any user-specified proposal and acceptance steps. 

\begin{table*}[t]

\setlength{\tabcolsep}{4.5pt}
\renewcommand{\arraystretch}{1.15}
\begin{ruledtabular}
\begin{tabular}{l l c c c c}
Reference & Method & State space & Move register & Ancilla & Total \\
\hline
Childs et al. \cite{childs2022quantum}&
Quantum walk for MALA&
$2pd$ & $0$ & $3pd+d_u$ & $5pd+d_u$ \\
Lemieux et al. \cite{lemieux2020efficient}&
Isometric circuit $\tilde{U}_W$ &
$m$ & $N$ & $N{+}2$ & $2N+m+2$ \\
This work (Theorem~\ref{the:summary})&
Dual kernel qubitized walk&
$4m$ & $0$ & $3$ & $4m+3$ \\
This work (Appendix~\ref{app:other}) &
Controlled-SWAP encoding &
$2m$ & $0$ & $1$ & $2m+1$ \\

\end{tabular}
\end{ruledtabular}
\caption{Logical qubit requirements for quantum circuit realizations of the Metropolis-Hastings algorithm.  \secondreview{Childs et al.~\cite{childs2022quantum} develop a quantum walk for a MALA where proposals are made following the underdamped Langevin process in continuous space. Considering that they directly encode each coordinate of $x\in \mathbb R^p$ in the qubit register with a precision of $d$ qubits, the total qubit count is $pd$ for one state space register $\ket{x}$. As their algorithm prepares $U\ket{x}\ket 0 = \ket{x}\int_\Omega \text{d}y \sqrt{p_{x\rightarrow y }}\ket{y}$, it requires two state space registers. Additionally, they use three other ancilla registers of the same size $pd$, and a register $\ket{u}$ in the range $[0,1]$, which we take to be discretized with $d_u$ qubits (see Algorithms 10 and 11 in~\cite{childs2022quantum}). In Lemieux et al.~\cite{lemieux2020efficient}, the circuits act on a system register of $m$ qubits, which represent spin configurations, and they encode the possible moves in unary in a Move register of $N$ qubits to enable a parallel implementation.
In our work, we consider a state space of size $n$ and use binary encoding with $m=\lceil \log_2 n\rceil$ qubits. Our main construction gives a qubitized walk operator on the dual space which requires four $m$-qubit state registers and three ancillae. The construction in Proposition~\ref{prop:2m-construction} achieves the cheapest costs with two $m$-qubit state registers and one ancilla.}}
\label{tab:costs}
\end{table*}

\section{Technical background}

\textbf{Notations.} Let $\mathbb S$ denote a finite set of size $n\in \mathbb N$, where $\mathbb N$ is the set of nonnegative integers. For any distribution $\mu:\mathbb S\to[0, 1]$ on $\mathbb S$, such that $\sum_{x\in \mathbb S}\mu(x)=1$, we denote its corresponding coherent state by $\ket\mu\in \mathbb C^{n}$ with $\braket{x|\mu}=\sqrt{\mu(x)}$ for each $x\in\mathbb S$. In particular, $\ket\pi=\sum_{x\in\mathbb S}\sqrt{\pi(x)}\ket x$ is the coherent state corresponding to the stationary distribution. Here, $\{\ket x\}_{x\in \mathbb S}$ denotes the canonical basis of $\mathbb C^n$. We may refer to both $\mu$ and $\ket\mu$ as the distribution. The presence or absence of the $\ket\cdot$ notation will make the actual subject clear. $\braket{\cdot, \cdot}$ corresponds to a standard Hermitian product: 
\begin{equation}
\forall f, g:\mathbb S\to\mathbb C:\braket{f, g}=\sum_{x\in\mathbb S}\overline{f(x)}g(x).
\end{equation}
Note that when an operator $H$ is symmetric, $H(x, y)=\overline{H(y, x)}$ for each $x, y\in\mathbb S$, we will adopt the Dirac notation:
\begin{equation}
\braket{f|H|g}=\braket{f, Hg}=\braket{Hf, g},
\end{equation}
for each $f,g:\mathbb S\to\mathbb C$. $1$ denotes the identity operator. The superscript $^\dagger$ denotes the conjugate transpose of an operator. For a linear operator $A$ acting on a finite-dimensional Hilbert space, $\sigma(A)$ is its set of eigenvalues. $\|\cdot\|$ refers to the spectral norm. $\ln$ is the natural logarithm, in base $e$. The \raisebox{-0.3em}{\includegraphics[height=1em]{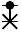}} circuit denotes the operation $\ket{x, y}\mapsto\ket{x, y\oplus x}$ for every bitstrings $x, y$ of the same length, where $\oplus$ is the bitwise addition modulo two. In other words, it is a tensor product of CNOT gates with controls in the first register and targets in the second register. $X=\ket0\bra1+\ket1\bra0=\ket+\bra+-\ket-\bra-$ and $Z=\ket0\bra0-\ket1\bra1$ are the usual Pauli operators. $H$ is the Hadamard matrix, so that $X=HZH$. For any $a, b\in \mathbb S$, $\delta_a(b)$ is $1$ if $a=b$ and $0$ otherwise.

The present section gathers well-known results related to Markov chains and quantum algorithms. A more comprehensive presentation of the Markov chain theory can be found in~\cite{levin2008markov}. Reference~\cite{sunderhauf2023generalizedquantumsingularvalue} contains more information about projected unitary encodings, qubitized walk operators and the Hermitianization procedure. The Szegedy quantization method is detailed in~\cite{1366222}.

\subsection{Ergodic Markov kernels and their mixing time}

\thirdreview{The central objects of our study are Markov kernels, that is, functions $P:\mathbb S^2\to[0,1]$ such that
\begin{equation}
\forall x\in \mathbb S:\sum_{y\in \mathbb S}P(x, y)=1.
\end{equation}}
\thirdreview{We will only consider Markov chains on finite state spaces that are ergodic. In particular, these chains have a single non-degenerate eigenvalue on the unit circle, and they are irreducible in the sense that
\begin{equation}\forall x, y\in\mathbb S: \exists t\in\mathbb N:P^t(x, y)>0.\end{equation}
}
For such kernels, $P^t$ converges as $t\to\infty$ to a probability distribution on $\mathbb S$: there exists a function $\pi:\mathbb S\to [0,1]$ such that
\begin{equation}
\forall x, y\in\mathbb S:\lim_{t\to\infty}P^t(x, y)=\pi(y).
\end{equation}
\thirdreview{We call $\pi$ the stationary distribution of $P$. The stationary distribution of an ergodic Markov chain is strictly positive element-wise: $\pi$ takes values in \firstreview{$(0, 1)$}. We define the spectral gap $\delta$ of an ergodic Markov kernel $P$ by
\begin{equation}
\delta=1-\max_{\lambda\in\sigma(P)\backslash\{1\}}|\lambda|.
\end{equation}}

\thirdreview{If a kernel admits a strictly positive left eigenvector, it is possible to define a scalar product induced by this eigenvector. The adjoint $P^\star$ of $P$ is its adjoint according to the scalar product induced by $\pi$. More precisely, for $f, g:\mathbb S\to\mathbb C$, define
\begin{equation}
\braket{f, g}_\pi=\sum_{x\in\mathbb S}\pi(x)\overline{f(x)}g(x).
\end{equation}
Then $P^\star$ is the unique operator such that:
\begin{equation}
\forall f, g:\mathbb S\to\mathbb C:\braket{f, Pg}_\pi=\braket{P^\star f, g}_\pi.
\end{equation}}

\thirdreview{One way to measure the distance between probability distributions is the total variation distance. For probability distributions $\mu$ and $\nu$ on $\mathbb S$, it is defined by
\begin{equation}
d_{TV}(\mu, \nu)=\frac12\sum_{x\in \mathbb S}|\mu(x)-\nu(x)|.
\end{equation}}

\thirdreview{For an ergodic chain $P$, convergence to the stationary distribution implies that for each probability distribution $\mu$ on $\mathbb S$, $d_{TV}(\mu P^t, \pi)\to0$ as $t\to\infty$. In particular, for $0<\epsilon <1$, the $\epsilon$-mixing time of $P$ is well-defined as
\begin{equation}
\tau(\epsilon)=\min \left\{t\in \mathbb N:\max_{x\in\mathbb S}d_{TV}\left(P^t(x, \cdot), \pi\right)\leq \epsilon\right\}.
\end{equation}}

\subsection{Reversible Markov chains}

\thirdreview{We say that a Markov kernel $P$ is reversible with respect to a probability measure $\pi$ on $\mathbb{S}$ if and only if
\begin{equation}
\forall x, y\in \mathbb S:\pi(x)P(x, y)=\pi(y)P(y, x).
\end{equation}
Equivalently, $P = P^\star$.}
Reversibility implies that $\pi$ is the stationary distribution, but the converse does not hold in general. 
For ergodic and reversible kernels, the relationship between the spectral gap and the mixing time is well-known~\cite{levin2008markov} and given by Theorem~\ref{the:gap_mix}.

\begin{The}
Let $P$ be an ergodic reversible Markov chain with stationary measure $\pi$ and $\epsilon\in (0, 1)$. Define $\pi_*=\min_{x\in \mathbb S}\pi(x)$. Then, the $\epsilon$-mixing time of $P$ satisfies:
\begin{equation}
\left(\frac{1}\delta-1\right)\ln\left(\frac1\epsilon\right)\leq \tau(\epsilon)\leq \frac1{\delta}\ln\left(\frac1{2\epsilon\sqrt{\pi_*}}\right).
\end{equation}
\label{the:gap_mix}
\end{The}

In particular, the mixing time is of the order of the inverse spectral gap.

\subsection{Projected unitary encodings and qubitized walk operators}

Let $\square:\mathbb C^j\to\mathbb C^l$, for $j, l\in \mathbb N$. Recall that $\square$ is a partial isometry if and only if $\square^\dagger\square$ is the projection on the support of $\square$. In this section, we will consider partial isometries and unitary operators acting on finite-dimensional Hilbert spaces. 

\thirdreview{Let $U$ be a unitary operator and $\square_L, \square_R$ be partial isometries. $(U, \square_L, \square_R)$ is said to be a Projected Unitary Encoding (PUE) of $A=\square_L^\dagger U\square_R$. If $\square_L=\square_R=\square$ and $U$ is also symmetric, $(U, \square)$ is called a Symmetric Projected Unitary Encoding (SPUE) of $A=\square^\dagger U\square$.}

\thirdreview{Let us now introduce qubitized walk operators. For a SPUE $(U, \square)$, its qubitized walk operator is
\begin{equation}
\mathcal W=(2\square\square^\dagger-1)U.
\end{equation}}
This closely resembles the Szegedy walk operators of the form $2\square\square^\dagger-1$, introduced in the next subsection. These two kinds of walk operators differ only by multiplication by a symmetric $U$ on the right-hand side. As stated by Theorem~\ref{the:qubitized_eigvals}, the added multiplication by $U$ ensures that qubitized walk operators leave (up-to-)two-dimensional subspaces invariant. Their dynamics is \firstreview{\textquotedblleft qubitized\textquotedblright}. 

\begin{The}
Let $(U, \square)$ be a SPUE of $A$. Let $\mathcal W$ be the qubitized walk operator of $(U, \square)$. If $\lambda \in ]-1, 1[$ is an eigenvalue of $A$ with eigenvector $\ket v$, then $e^{\pm i\cos^{-1}(\lambda)}$ is an eigenvalue of $\mathcal W$ with normalized eigenvectors:
\begin{equation}
\ket{\mu(\lambda)^{\pm}}=\frac{1}{\sqrt{2}\sin(\theta)}\left(e^{\pm i\theta}-U\right)\square\ket{v},
\label{eq:eigvecs}
\end{equation}
where $\theta=\cos^{-1}(\lambda)$. If $\pm1$ is an eigenvalue of $A$ with eigenvector $\ket v$, then $\square\ket v$ is an eigenvector of $\mathcal W$ with the same eigenvalue. Moreover, it is a complete description of the spectrum of $\mathcal W$ in the sense that:
\begin{equation}
\sigma\left(\mathcal W\right)\backslash\{\pm1\}=\left\{e^{\pm i\cos^{-1}(\lambda)}:\lambda\in \sigma(A)\cap]-1, 1[\right\},
\end{equation}
with multiplicities.
\label{the:qubitized_eigvals}
\end{The}

Because of Theorem~\ref{the:qubitized_eigvals}, SPUEs play a central role in the design of quantum algorithms~\cite{sunderhauf2023generalizedquantumsingularvalue}. Starting from a PUE, we will often work with its so-called Hermitianization, its symmetric counterpart defined in the following Proposition~\ref{prop:hermitianization} (see for example~\cite{sunderhauf2023generalizedquantumsingularvalue}).

\begin{Prop} Let $\left(U, \square_L, \square_R\right)$ be a PUE of $A$. Define $\overline U=\left(\ket0\bra0\otimes U+\ket1\bra1\otimes U^\dagger\right)\left(X\otimes 1\right)$ and $\overline \square=\ket0\bra0\otimes \square_L+\ket1\bra1\otimes\square_R$. Then, $\left(\overline U, \overline \square\right)$ is a SPUE of $\overline A=\ket0\bra1\otimes A+\ket1\bra0\otimes A^\dagger$. In particular, $\overline U$ can be constructed from a controlled-$U$ gate, its adjoint and an $X$ gate. $\left(\overline U, \overline \square\right)$ is called the Hermitianization of $(U, \square_L, \square_R)$. Moreover, the eigenvalues of $\overline A$ are plus or minus the singular values of $A$.
\label{prop:hermitianization}
\end{Prop}

\subsection{Szegedy quantization of Markov chains}

\firstreview{Let us now introduce quantum step operators. Recall that for every state $x\in \mathbb S$, the corresponding row $P(x, \cdot)$ of the Markov kernel is a probability distribution. In particular, we denote its coherent encoding by 
\begin{equation}
\ket{P(x, \cdot)}=\sum_{y\in\mathbb S}\sqrt{P(x, y)}\ket y.
\end{equation}}

\thirdreview{Given a Markov kernel $P$, its quantum step operator $\square:\mathbb C^n\to\mathbb C^{n^2}$ is
\begin{equation}
\square = \sum_{x\in \mathbb S}\ket{x, P(x, \cdot)}\bra x.
\end{equation}
Furthermore, the Szegedy walk operator $R:\mathbb C^{n^2}\to\mathbb C^{n^2}$ of $P$ is
\begin{equation}
R = 2\square\square^\dagger-1.
\end{equation}}

We note that the Szegedy walk operator is the qubitized walk operator of $(1, \square)$. Indeed,
\begin{equation}
\begin{split}
\square^\dagger\square&=\left(\sum_{x\in \mathbb S}\ket x\bra{x, P(x, \cdot)}\right)\left(\sum_{y\in \mathbb S}\ket{y, P(y, \cdot)}\bra y\right)\\
&=\sum_{x\in \mathbb S}\ket x\bra x,
\end{split}
\end{equation}
and $\square$ is a partial isometry. Moreover, if $(O, 1, \ket0)$ is a PUE of the quantum step operator $\square$, then we can construct the unitary $R$ using the equation:
\begin{equation}
R = 2O\ket0\bra0O^\dagger-1=O\left(2\ket0\bra0-1\right)O^\dagger.
\label{eq:reflection_quantumstep}
\end{equation}
Since reflecting through the $\ket0$ state is an easy task, it is sufficient to construct such PUEs of the quantum step operator to implement the Szegedy walk operator efficiently. Let $S:\mathbb C^{n^2}\to\mathbb C^{n^2}$ be the SWAP operator $S\ket{x, y}=\ket{y, x}$.

\begin{Prop} Let $P$ be a reversible ergodic kernel with stationary measure $\pi$ and spectral gap $\delta$. Then, $(S, \square)$ is a SPUE of $D=\square^\dagger S\square$ where the discriminant $D$ of $P$ is:
\begin{equation}
\forall x, y\in \mathbb S: D(x, y)=\sqrt{\frac{\pi(x)}{\pi(y)}}P(x, y).
\label{def:discriminant}
\end{equation}
In particular, $\square\ket\pi$ is the only eigenvector of $\mathcal W$, the qubitized walk operator of $(S, \square)$, in the range of $\square$. Furthermore, $\mathcal W$ has a quadratically amplified angular gap $\Delta$ compared to $P$: 
\begin{equation}
\begin{split}
\Delta&=\min\left\{\theta\neq 0:e^{i\theta}\in \sigma(\mathcal W)\right\}\\
&=\cos^{-1}\left(\max\{\sigma(P)\backslash\{1\}\}\right)\\
&\in\Omega\left(\sqrt{\delta}\right).
\end{split}
\end{equation}
\label{prop:quadratic_speedup_for_reversible}
\end{Prop}

In particular, Proposition~\ref{prop:quadratic_speedup_for_reversible} states that the spectral gap of the qubitized walk operator is quadratically larger than that of the Markov process.

\subsection{Metropolis-Hastings algorithm}

\begin{figure}[h]
    \centering
    \includegraphics[width=1\linewidth]{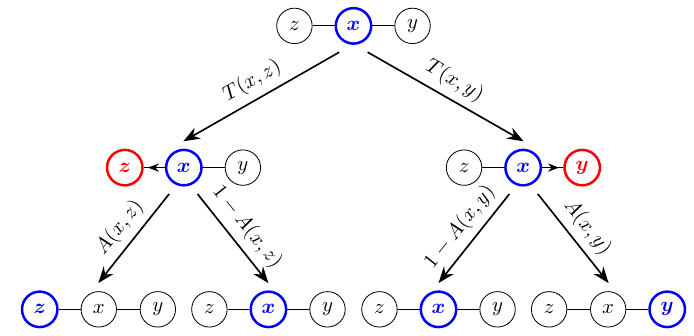}
    \caption{Graphical representation of a MH step. Step 1. The process is in the state $x$, with neighbours $y, z$. Step 2. Select the outgoing edge $(x, y)$ with probability $T(x, y)$ and the outgoing edge $(x, z)$ with probability $T(x, z)$. Step 3. Given the current edge $(a, b)$, let the process be in state $b$ with probability $A(a, b)$ and be in state $a$ with probability $1-A(a, b)$.}
    \label{fig:MH}
\end{figure}

The Metropolis-Hastings algorithm~\cite{10.1093/biomet/57.1.97} aims at sampling from a probability distribution $\pi:\mathbb S\to(0, 1)$ while only being able to compute ratios of equilibrium probabilities $\pi(x)/\pi(y)$ for all $x, y\in\mathbb S$. It consists in designing a reversible ergodic Markov kernel $P$ from a proposal kernel $T$ and move acceptance probabilities $A$. Starting from a state $x\in \mathbb S$, propose a new state $y\in \mathbb S\backslash\{x\}$ according to the distribution $T(x, \cdot)$. Then, accept this new state with probability $A(x, y)$. Otherwise, stay in state $x$. Figure~\ref{fig:MH} illustrates the proposal-acceptance MH steps from a state with two neighbours.

\thirdreview{A Markov kernel $T$ on $\mathbb S$ is said to be a valid proposal kernel if $T(x, y)>0\implies T(y, x)>0$ for every $x\neq y\in \mathbb S$ and $T(x, x)=0$ for each $x\in \mathbb S$. Let $A:\mathbb S^2\to(0, 1]$ be the so-called acceptance probability matrix. The Metropolis-Hastings kernel $P$ associated with $T$ and $A$ is defined by
\begin{equation}
P(x, y)=\left\{
\begin{aligned}
&T(x, y)A(x, y) \text{ if $x\neq y$,}\\
&1-\sum_{z\in \mathbb S}T(x, z)A(x, z) \text{ if $x=y$,}
\end{aligned}
\right.
\label{eq:mhkernel}
\end{equation}
for all $x, y\in \mathbb S$.}

The reversibility condition can be written:
\begin{equation}
\pi(x)T(x, y)A(x, y)=\pi(y)T(y, x)A(y, x),
\label{eq:reversibility_condition}
\end{equation}
for all $x, y\in \mathbb S$ and can be satisfied by carefully choosing the matrix $A$.

\thirdreview{Two standard choices of $A$ satisfy this condition. Let $\pi:\mathbb S\to(0, 1)$ be a probability distribution on $\mathbb S$ and let $T$ be a valid proposal kernel. The Metropolis-Hastings kernel $P$ defined from $T$ and $A$ is said to use the Glauber choice when, for all $x, y\in\mathbb S$ such that $T(x, y)>0$,
\begin{equation}
A(x, y)=\frac{1}{1+\frac{\pi(x)T(x, y)}{\pi(y)T(y, x)}},
\label{eq:glauber_choice}
\end{equation}
and the Metropolis choice when, for the same pairs $x,y$,
\begin{equation}
A(x, y)=\min\left(1, \frac{\pi(y)T(y, x)}{\pi(x)T(x, y)}\right),
\label{eq:metropolis_choice}
\end{equation}}

If $P$ is a Metropolis-Hastings kernel with Glauber or Metropolis choice, then Equation~\ref{eq:reversibility_condition} is clearly satisfied, i.e. $P$ is reversible with respect to $\pi$. 


\subsection{The quantum proposal and acceptance operators}

Let $P$ be an ergodic and reversible Metropolis-Hastings kernel with stationary measure $\pi$. The classical simulation of $P$ consists in sampling from a proposal kernel $T$ and the computation of the matrix elements of an acceptance probability matrix $A:\mathbb S^2\to[0, 1]$. Hence, we only want to rely on the simulation of the proposal kernel, using a PUE $(O_T, 1, \ket0)$ of $\square_T$ defined by:
\begin{equation}
O_T\ket0=\square_T = \sum_{x\in \mathbb S}\ket{x, T(x, \cdot)}\bra x,
\end{equation}
and a PUE $(O_A, 1, \ket0)$ such that:
\begin{equation}
\braket{x, y|O_A|x, y}=\begin{pmatrix}
\sqrt{1-A(x, y)}    & -\sqrt{A(x, y)}\\
\sqrt{A(x, y)}      & \sqrt{1-A(x, y)}
\end{pmatrix},
\label{eq:def_OA}
\end{equation}
for every $x, y\in \mathbb S$. Note that, in particular, 
\begin{equation}
\begin{split}
&O_A\ket{x, y}\ket0\\
&=\sqrt{1-A(x, y)}\ket{x, y}\ket0+\sqrt{A(x, y)}\ket{x, y}\ket1.
\end{split}
\end{equation}
for every $x, y\in \mathbb S$. \secondreview{We emphasize that our framework assumes access to $O_A$, and that the cost of realizing it is application dependent. Therefore, our contribution is that, given $O_T$ and $O_A$, the quantum walk construction does not require further coherent arithmetic routines to evaluate the rejection probabilities $P(x, x)$. In Appendix~\ref{app:OAOT}, we give an explicit single-ancilla implementation of $O_A$ in the regime where $\theta(x, y)$ approximates a smooth function.}

\section{Results}
\thirdreview{We now give the mathematical results behind our construction. This section is organized from the classical dual kernel to the final quantum circuit: we first define the edge-space chain, then construct quantum step operators for the dual kernel and its time reversal. Since the dual kernel is generally nonreversible, we construct its reversible dilation before applying the qubitized-walk spectral theorem, Theorem~\ref{the:qubitized_eigvals}. We finally prove that the resulting qubitized walk keeps the quadratic gap amplification, and show how to use the algorithm to extract samples of $\pi$.}
\subsection{Problem statement}

Let $P$ be an ergodic and reversible Metropolis-Hastings kernel with stationary measure $\pi$. Using solely $O_T$ and $O_A$, our goal is to construct the quantum update step of a reversible Markov kernel with stationary measure $\pi$ and the same spectral gap as $P$. We will construct the quantum step operator of a Markov kernel on $\mathbb S^2$ with stationary measure $\pi(x)T(x, y)$. The first marginal of this distribution is precisely $\pi$. Moreover, this kernel has exactly the same mixing time as $P$ (under the Glauber choice). Hence, its spectral gap is of the same order of magnitude as that of $P$ and the quadratic speedup is maintained. Since it does not require performing arithmetic on qubit registers, our construction significantly reduces the resource requirements of the MH Szegedy quantization. 

\subsection{Dual Metropolis-Hastings kernel}\label{sec:duals}

In this subsection, we introduce Markov processes on $\mathbb S^2$, the set of pairs of states. The enlarged state space allows to keep track of rejected samples, facilitating their implementation through invertible unitary operations. For every pair $(x, y)\in \mathbb S^2$, we will refer to $x$ as the tail and $y$ as the head. Denote by $\mathcal S\subset\mathbb S^2$ the set of edges of the proposal kernel's graph:
\begin{equation}
\mathcal S=\{(x, y)\in\mathbb S^2:T(x, y)>0\}.
\end{equation}
For any probability measure $F$ on the edges in $\mathcal S$, we will refer to the probability distribution of the tails as the first marginal $\mu$ and the probability distribution of the heads given the tail as the conditional distribution $Q$. In particular, $F(x, y)=\mu(x)Q(x, y)$ for every $(x, y)\in\mathcal S$, $\mu$ is a probability distribution on $\mathbb S$ and $Q$ is a Markov kernel on $\mathbb S$. 

\firstreview{Before presenting our construction, we first describe the dual MH kernel and why it is preferable to the usual MH kernel. In the MH algorithm, given the current state $x\in \mathbb S$, we propose a new state $y$ according to the proposal kernel. Then, we either accept this state, in which case the new state is $y$, or we reject it, in which case the new state is $x$. The difficulty in implementing the MH algorithm quantumly comes from the need to replace any $y$ by $x$ whenever $y$ is rejected, since such an operation is non-invertible. The dual kernel reformulates the chain such that rejection operations become invertible. We consider a state $(x, z)\in\mathcal S$ and propose the state $(x, y)$ with probability $T(x, y)$. The acceptance now acts on edges, rather than on individual states: with probability $A(x,y)$, we accept $y$ and flip the edge to $(y, x)$; with probability $1-A(x,y)$, the proposal is rejected and we remain at $(x,z)$. Both flipping the edge and doing nothing are invertible operations, so this process can be embedded in unitary operations. Moreover, the tail of the edge evolves according to a Markov process whose transition probabilities are given by the MH kernel, since the proposal does not depend on the heads. Consequently, the tail marginal of the process tends to $\pi$, as in the standard MH algorithm.}

\subsubsection{Dual proposal $\mathcal T$ and acceptance $\mathcal A$ kernels}

Let us introduce the dual proposal kernel on $\mathbb S^2$. Given a pair of states in $\mathbb S$, the dual proposal kernel proposes a new head while keeping the same tail.
\begin{Def} Define the Markov kernel $\mathcal T$ on $\mathbb S^2$ by:
\begin{equation}
\forall (x, y), (z, t)\in \mathbb S^2:\mathcal T((x, y), (z, t))=\delta_x(z)T(x, t).
\end{equation}
\end{Def}
Sampling a new edge from $\mathcal T$ always results in an edge in $\mathcal S$. Therefore, our subsequent analysis can safely be reduced to $\mathcal S$. Given a probability distribution on $\mathcal S$, $\mathcal T$ returns a probability distribution with the same first marginal but with conditional distribution replaced by $T$. 
\begin{Lem}
Let $F:\mathcal S\to[0, 1]$ be a probability measure with first marginal $\mu$ and conditional $Q$. Then, $F\mathcal T$ has first marginal $\mu$ and conditional $T$.
\label{lem:action_of_T}
\end{Lem}
\begin{proof}
Let $(x, y)\in \mathcal S$. Then, 
\begin{equation}
\begin{split}
F\mathcal T(x, y)&=\sum_{(z, t)\in \mathcal S}\mu(z)Q(z, t)\delta_x(z)T(x, y)\\
&=\sum_{z\in \mathbb S}\mu(z)\delta_x(z)T(x, y)\\
&=\mu(x)T(x, y).
\end{split}
\end{equation}
\end{proof}
\thirdreview{We introduce a key probability measure $\nu:\mathbb S^2\to[0, 1]$ given by
\begin{equation}
\nu(x, y)=\pi(x)T(x, y),\qquad \forall x,y\in \mathbb S.
\end{equation}
Its support is $\mathcal S$ and its first marginal is $\pi$.}
Whereas the proposal kernel $T$ is not necessarily reversible with respect to any measure, its dual $\mathcal T$ is reversible with respect to $\nu$. From now on, the $^\star$ superscript will always denote the adjoint with respect to $\langle\cdot, \cdot\rangle_\nu$.

\begin{Prop} $\mathcal T$ is reversible with respect to $\nu$.
\label{prop:dualTreversible}
\end{Prop}

\begin{proof}
Let $(x, y), (z, t)\in \mathbb S^2$. Then,
\begin{equation}
\begin{split}
\nu(x, y)\mathcal T((x, y), (z, t))&=\pi(x)T(x, y)\delta_{x}(z)T(x, t)\\
&=\pi(z)T(z, y)\delta_z(x)T(z, t)\\
&=\nu(z, t)\delta_z(x)T(z, y)\\
&=\nu(z, t)\mathcal T((z, t), (x, y)).
\end{split}
\end{equation}
\end{proof}
Let us now define the dual $\mathcal A$ of the acceptance matrix. $\mathcal A$ flips the input edge $(x, y)\in \mathbb S^2$ to $(y, x)$ with probability given by the acceptance matrix $A$. Otherwise, the edge remains in state $(x, y)$. Compared to the MH algorithm, the rejected sample is not completely forgotten but stored in the head.

\begin{Def} Define the Markov kernel $\mathcal A$ on $\mathbb S^2$ by:
\begin{equation}
\forall x, y\in \mathbb S:
\left\{
\begin{aligned}
&\mathcal A((x, y), (y, x))=A(x, y)\\
&\mathcal A((x, y), (x, y))=1-A(x, y).
\end{aligned}
\right.
\end{equation}
\end{Def}

The reversibility condition (Eq.~\ref{eq:reversibility_condition}) implies that $\mathcal A$ has to be reversible with respect to $\nu$.

\begin{Prop} If $P$ is a reversible Markov kernel, then $\mathcal A$ is reversible with respect to $\nu$.
\label{prop:dualAreversible}
\end{Prop}

\begin{proof} Let $x\neq y\in \mathbb S$. Then, 
\begin{equation}
\begin{split}
\nu(x, y)\mathcal A((x, y), (y, x))&=\pi(x)T(x, y)A(x, y)\\
&=\pi(x)P(x, y)\\
&=\pi(y)P(y, x)\\
&=\pi(y)T(y, x)A(y, x)\\
&=\nu(y, x)\mathcal A((y, x), (x, y)),
\end{split}
\end{equation}
using the reversibility condition of $P$, Equation~\ref{eq:reversibility_condition}.
\end{proof}

We can now define the dual Metropolis-Hastings kernel $\mathcal P$ of $P$.
\begin{Def}
Define the dual Metropolis-Hastings kernel $\mathcal P$ on $\mathbb S^2$ by the equation $\mathcal P=\mathcal{TA}$.
\end{Def}
$\mathcal P$ acts on the first marginals as $P$, as stated precisely by Lemma~\ref{lem:action_of_P}.

\begin{Lem} Let $F:\mathcal S\to[0, 1]$ be a probability measure with first marginal $\mu$. Then, $F\mathcal P$ has first marginal $\mu P$.
\label{lem:action_of_P}
\end{Lem}

\begin{proof}
Let $(x, y)\in \mathcal S$. By Lemma~\ref{lem:action_of_T}, $F\mathcal T(x, y)=\mu(x)T(x, y)$ for every $(x, y)\in \mathcal S$. Therefore, 
\begin{equation}
\begin{split}
&F\mathcal T\mathcal A(x, y)\\
&=\mu(x)T(x, y)(1-A(x, y))+\mu(y)T(y, x)A(y, x).
\end{split}
\end{equation}
Computing the first marginal,
\begin{equation}
\begin{split}
&\sum_{y:(x, y)\in \mathcal S}F\mathcal T\mathcal A(x, y)\\
&=\sum_{y:(x, y)\in \mathcal S}\mu(x)T(x, y)(1-A(x, y))\\
&+\mu(y)T(y, x)A(y, x)\\
&=\mu(x)P(x, x)+\sum_{y\in \mathbb S\backslash\{x\}}\mu(y)P(y, x)\\
&=\mu P(x).
\end{split}
\end{equation}

\end{proof}

\begin{figure}[h]
    \centering
    \includegraphics[width=1\linewidth]{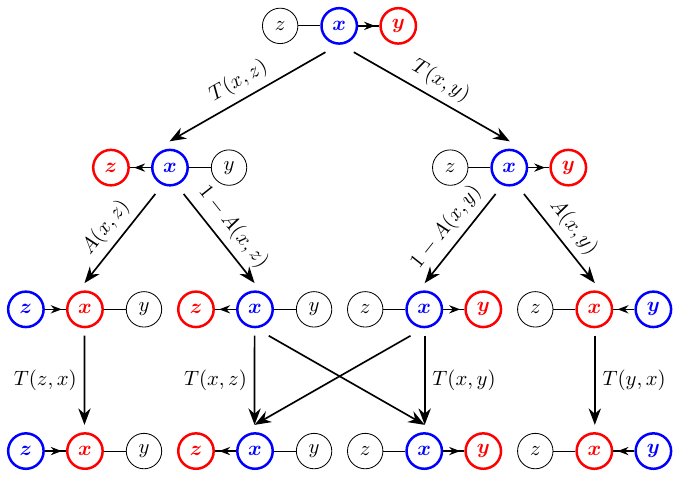}
    \caption{Graphical representation of a $\mathcal{TAT}$ step. Step 1. The process is in the state $(x, y)$. Step 2. Select the new head $y$ with probability $T(x, y)$ and the new head $z$ with probability $T(x, z)$. Step 3. Given the current edge $(a, b)$, flip it to $(b, a)$ with probability $A(a, b)$. Step 4. Sample a new head for the current edge as in Step 2.}
    \label{fig:TAT}
\end{figure}

Since $\mathcal P$ is the product of two reversible kernels, $\nu$ is a stationary measure of $\mathcal P$. However, $\mathcal P$ is not necessarily reversible with respect to $\nu$. Still, a step of $\mathcal P$ followed by a step of $\mathcal T$ results in a reversible process. Proposition~\ref{prop:structure_of_TAT} describes the action of $\mathcal{TAT} = \mathcal{PT}$ on the marginal and conditional distributions. Figure~\ref{fig:TAT} illustrates how $\mathcal{TAT}$ selects the next edge.

\begin{Prop}
For every probability distribution $F$ on $\mathcal S$ with first marginal $\mu$, 
\begin{equation}
\forall (x, y)\in \mathcal S:F\mathcal{TAT}(x, y)=(\mu P)(x)T(x, y).
\end{equation}
In words, $F\mathcal{TAT}$ has first marginal $\mu P$ and conditional $T$.
\label{prop:structure_of_TAT}
\end{Prop}

\begin{proof}
By Lemma~\ref{lem:action_of_P}, $\mathcal{TA}=\mathcal P$ applies $P$ to the first marginal. By Lemma~\ref{lem:action_of_T}, $\mathcal T$ takes any conditional distribution to $T$ and does not affect the first marginal.
\end{proof}

Proposition~\ref{prop:structure_of_TAT} directly implies that $\mathcal{TAT}$ is ergodic (on its essential states $\mathcal S$) and has the same mixing times as $P$. In particular, its spectral gap must be of the same order as that of $P$. As it turns out, we can deduce a lot more from Proposition~\ref{prop:structure_of_TAT}.

\begin{Cor}
Let us consider the restriction of the dual kernel to $\mathcal S$. The Markov kernel $\mathcal{TAT}$ is reversible with respect to $\nu$. Moreover, $\mathcal{TAT}$ and $P$ have the same spectrum, up to trivial eigenvalues: 
\begin{equation}
\sigma(P)\subset\sigma(\mathcal{TAT})\subset\sigma(P)\cup\{0\}.
\end{equation}
In particular, $\mathcal{TAT}$ is ergodic. Note that considering $\mathcal{TAT}$ on $\mathbb S^2$ only adds $0$ eigenvalues.
\label{cor:equal_mixing}
\end{Cor}

\begin{proof}
The first claim follows from $\mathcal T=\mathcal T^\star$ (Proposition~\ref{prop:dualTreversible}) and $\mathcal A=\mathcal A^\star$ (Proposition~\ref{prop:dualAreversible}). Now let $\lambda\in \sigma(\mathcal{TAT})$. Let $f:\mathcal S\to[-1, 1]$ be a real left eigenvector of $\mathcal{TAT}$. Decompose $f$ as the sum of its positive and negative parts: $f=f_{+}-f_-$, where $f_{\pm}:\mathcal S\to[0, 1]$. Without loss of generality, $f$ can be normalized such that $f_++f_-$ is a probability measure on $\mathcal S$. For every $x\in \mathbb S$, define
\begin{equation}
\mu_{\pm}(x)=\sum_{y:(x, y)\in \mathcal S}f_{\pm}(x, y).
\end{equation}
Then, \thirdreview{still without loss of generality},
\begin{equation}
f_{\pm}(x, y)=\mu_{\pm}(x)Q_{\pm}(x, y),
\end{equation}
for some Markov kernels $Q_{\pm}$ on $\mathbb S$. With these notations,
\begin{equation}
f(x, y)=\mu_+(x)Q_+(x, y)-\mu_-(x)Q_-(x, y).
\end{equation}
Using Proposition~\ref{prop:structure_of_TAT}, the eigenvalue equation can be written as:
\begin{equation}
\lambda f(x, y)=((\mu_+P)(x)-(\mu_-P)(x))T(x, y),
\end{equation}
for every $(x, y)\in \mathcal S$. \thirdreview{At this stage, there are two possibilities: either $\mu_+=\mu_-$ or $\mu_+\neq\mu_-$. If $\mu_+=\mu_-$, then $\lambda f=0$ and $\lambda=0$. If $\mu_+\neq\mu_-$,} summing over all $y\in \mathbb S$,
\begin{equation}
\lambda (\mu_+-\mu_-)=(\mu_+-\mu_-)P,
\end{equation}
namely $\lambda \in \sigma(P)$. Therefore, $\sigma(\mathcal{TAT})\subset\sigma(P)\cup\{0\}$. Let now $\lambda\in \sigma(P)$. Define $f:\mathcal S\to[-1, 1]$ by:
\begin{equation}
f(x, y)=g(x)T(x,y),
\end{equation}
where $g$ is a left eigenvector of $P$ with eigenvalue $\lambda$. Decompose $g=\mu_+-\mu_-$ with $\mu_{\pm}:\mathbb S\to[0, 1]$. Using Proposition~\ref{prop:structure_of_TAT},
\begin{equation}
\begin{split}
f\mathcal{TAT}(x, y)&=((\mu_+P)(x)-(\mu_-P)(x))T(x, y)\\
&=\lambda f(x, y),
\end{split}
\end{equation}
for every $(x, y)\in \mathcal S$. Therefore $\sigma(P)\subset\sigma(\mathcal{TAT})$.

\end{proof}

Any SPUE of the discriminant of $\mathcal{TAT}$ would therefore have a qubitized walk operator of angular gap $\Omega\left(\sqrt\delta\right)$. We can easily construct such a SPUE from the step operators of $\mathcal T$ and $\mathcal A$ with $4$ edge registers. In what follows, we provide a construction of the same efficiency with only $2$ edge registers.

\subsection{Quantum circuits for the dual quantum walk}\label{sec:qstepops}

In this subsection, we describe the encoding of the quantum step operators of $\mathcal P=\mathcal{TA}$ and $\mathcal P^\star=\mathcal{AT}$. Let us start with a technical modification of $O_A$. Even though we are mainly interested in its action on states $\ket{x, y, x}$ for $(x, y)\in \mathcal S$, we make sure that it preserves the $\ket0$ state of the ancilla for every possible input state.

\begin{Lem}
\thirdreview{We can implement a circuit $O_{\mathcal A}$ such that for every $x\neq y$, $\braket{0|O_{\mathcal A}|0}$ maps:
\begin{equation}
\left\{
\begin{aligned}
\ket{x, y, x}&\to\sqrt{1-A(x, y)}\ket{x, x, y}+\sqrt{A(x, y)}\ket{x, y, x},\\
\ket{x, x, y}&\to-\sqrt{A(x, y)}\ket{x, x, y}+\sqrt{1-A(x, y)}\ket{x, y, x}.
\end{aligned}
\right.
\end{equation}
Moreover, $\ket{x, y, z}$ is a fixed point of $\braket{0|O_{\mathcal A}|0}$ whenever $x, y$, and $z$ are all pairwise different.}
\label{lem:OmathcalA}
\end{Lem}

\begin{proof}
Impose that $A(x, x)=0$ for every $x\in \mathbb S$. Distinguish the possible input states:
\begin{equation}
\ket{x, y, z}\ket0=\left\{
\begin{aligned}
&\ket{x, y, x}\ket0,\\
&\ket{x, x, z}\ket0,\\
&\ket{x, y, z}\ket0,\\
&\ket{x, x, x}\ket0,
\end{aligned}
\right.
\end{equation}
with $x, y, z\in \mathbb S$ all pairwise different. Flip the ancilla if the first two registers are in the same state:
\begin{equation}
\to\left\{
\begin{aligned}
&\ket{x, y, x}\ket0,\\
&\ket{x, x, z}\ket1,\\
&\ket{x, y, z}\ket0,\\
&\ket{x, x, x}\ket1.
\end{aligned}
\right.
\end{equation}

To perform this operation, the first register is copied to the second by a succession of bitwise CNOT gates. If $x = y$, the second register will be in the state $\ket{x\oplus x} =\ket{0}$. A multi-controlled Toffoli gate, with the second register being in state $\ket{0}$ as control and the ancilla as target, achieves the desired result. Finally, we uncompute the copy operation with another succession of bitwise CNOT gates from the first register to the second.

If the ancilla is in state $\ket1$, swap the second and third registers, 
\begin{equation}
\to\left\{
\begin{aligned}
&\ket{x, y, x}\ket0,\\
&\ket{x, z, x}\ket1,\\
&\ket{x, y, z}\ket0,\\
&\ket{x, x, x}\ket1.
\end{aligned}
\right.
\end{equation}

Now apply a controlled-$O_A$ gate, controlled by the third register being equal to the first, which leads to:

\begin{equation}
\to\left\{
\begin{aligned}
&\sqrt{1-A(x, y)}\ket{x, y, x}\ket0+\sqrt{A(x, y)}\ket{x, y, x}\ket1,\\
&-\sqrt{A(x, z)}\ket{x, z, x}\ket0+\sqrt{1-A(x, z)}\ket{x, z, x}\ket1,\\
&\ket{x, y, z}\ket0,\\
&\ket{x, x, x}\ket1.
\end{aligned}
\right.
\end{equation}

Since $A(x,x) = 0$, the gate has no effect on the state $\ket{x,x,x}\ket{1}$. If the ancilla is in state $\ket{0}$, swap the second and third registers:
\begin{equation}
\to\left\{
\begin{aligned}
&\sqrt{1-A(x, y)}\ket{x, x, y}\ket0+\sqrt{A(x, y)}\ket{x, y, x}\ket1,\\
&-\sqrt{A(x, z)}\ket{x, x, z}\ket0+\sqrt{1-A(x, z)}\ket{x, z, x}\ket1,\\
&\ket{x, z, y}\ket0,\\
&\ket{x, x, x}\ket1.
\end{aligned}
\right.
\end{equation}
If the first and third registers are equal, flip the ancilla: 
\begin{equation}
\to\left\{
\begin{aligned}
&\sqrt{1-A(x, y)}\ket{x, x, y}\ket0+\sqrt{A(x, y)}\ket{x, y, x}\ket0,\\
&-\sqrt{A(x, z)}\ket{x, x, z}\ket0+\sqrt{1-A(x, z)}\ket{x, z, x}\ket0,\\
&\ket{x, z, y}\ket0,\\
&\ket{x, x, x}\ket0.
\end{aligned}
\right.
\end{equation}
\end{proof}

\begin{figure}
    \centering
    \includegraphics[width=\linewidth]{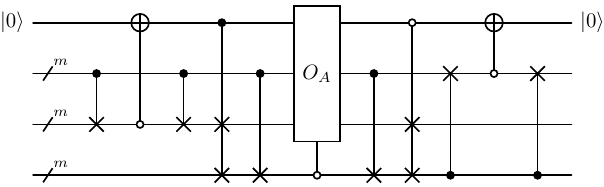}
    \caption{\thirdreview{Quantum circuit for $O_{\mathcal A}$, realizing the map described in Lemma~\ref{lem:OmathcalA}. The circuit makes use of a controlled-$O_A$ unitary.}}
    \label{fig:OmathcalA}
\end{figure}

The resulting quantum circuit is shown in Figure~\ref{fig:OmathcalA}. We are now ready to encode the step operator of $\mathcal P$.

\begin{Prop} We can construct a PUE $(O, \ket{0}, \ket{0}\ket0)$ of the quantum step operator $\square$ associated with the kernel $\mathcal P$. The construction of $O$ uses $O_{T}$ and $O_{\mathcal A}$ once.
\label{prop:construct_dual}
\end{Prop}

\begin{proof}
Let $(x, y)\in \mathcal S$. Using $O_{T}$ and a copy of the first register in the fourth, we can implement:
\begin{equation}
\ket{x, y}\to \ket{x, y, T(x, \cdot), x}=\sum_{z:(x,z)\in \mathcal S}\sqrt{T(x, z)}\ket{x, y, z, x}.
\end{equation}
Using $O_{\mathcal A}$ (Lemma~\ref{lem:OmathcalA}) on the first, third, and fourth registers:
\begin{equation}
\begin{split}
&\ket{x, y, T(x, \cdot), x}\\&\to \sum_{z:(x,z)\in \mathcal S}\sqrt{T(x, z)(1-A(x, z))}\ket{x, y, x, z}\\&+\sum_{z:(x,z)\in \mathcal S}\sqrt{T(x, z)A(x, z)}\ket{x, y, z, x}\\
&=\ket{x, y, \mathcal P((x, y), \cdot))}.
\end{split}
\end{equation}
\end{proof}

The quantum step operator associated with $\mathcal P^\star$ can also be implemented efficiently.

\begin{Lem}
We can implement a circuit $O_{\mathcal A\star}$ such that:
\begin{equation}
\begin{split}
&\forall (z,t)\in \mathcal S: \bra{0}(O_{\mathcal A\star}\ket{0})\ket{z, t, z}\\
&=\sqrt{1-A(z, t)}\ket{z, t, z}+\sqrt{A(z, t)}\ket{z, t, t}.
\end{split}
\end{equation}
\label{lem:OmathcalAstar}
\end{Lem}

\begin{proof}
$O_{\mathcal A\star}$ can be obtained by applying $O_{\mathcal A}$ followed by the transformation:
\begin{equation}
\ket{x, y, z}\to\ket{x, z, y}\to\ket{x, x\oplus y\oplus z, y}.
\end{equation}
Indeed, this transformation maps $\ket{z, z, t}$ to $\ket{z, t, z}$ and $\ket{z, t, z}$ to $\ket{z, t, t}$ for every $z\neq t\in \mathbb S$.
\end{proof}

\begin{Prop} We can construct a PUE $(O_\star, \ket{0}, \ket{0}\ket0)$ of the quantum step operator $\square_\star$ of $\mathcal P^\star$ with a single use of $O_{\mathcal A\star}$ and $O_T$.
\label{prop:construct_dual_star}
\end{Prop}

\begin{proof}
Let $(z, t)\in \mathcal S$. Copying $z$ to the third register and applying $O_{\mathcal A\star}$ (Lemma~\ref{lem:OmathcalAstar}), we obtain
\begin{equation}
\begin{split}
&\ket{z, t}\to \left(\sqrt{A(z, t)}\ket{z, t, t}+\sqrt{1-A(z, t)}\ket{z, t, z}\right)\ket0.
\end{split}
\end{equation}
Using $O_T$ to propose from the third register,
\begin{equation}
\begin{split}
&\left(\sqrt{A(z, t)}\ket{z, t, t}+\sqrt{1-A(z, t)}\ket{z, t, z}\right)\ket0\\
&\to \sum_{y:(t, y)\in \mathcal S}\sqrt{A(z, t)T(t, y)}\ket{z, t, t, y}\\
&+\sum_{y:(z, y)\in \mathcal S}\sqrt{(1-A(z, t))T(z, y)}\ket{z, t, z, y}\\
&=\ket{z, t, \mathcal P^\star((z, t), \cdot)}.
\end{split}
\end{equation}

\end{proof}

\begin{figure*}
    \centering
    \includegraphics[width=\linewidth]{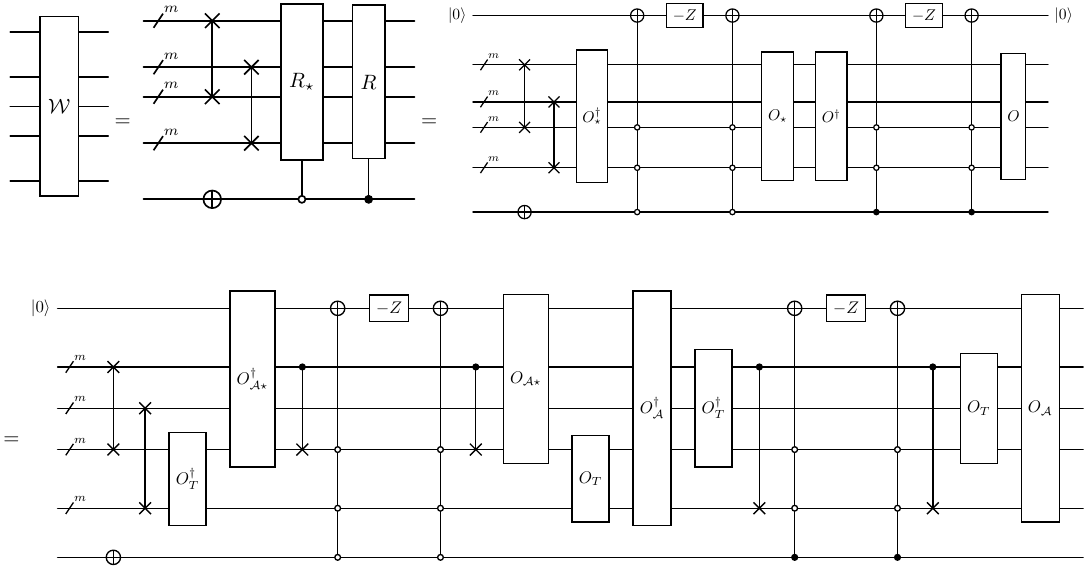}
    \caption{Quantum circuit for the final qubitized walk operator $\mathcal W$. Recall that we are quantizing a walk on the edges $\mathcal S$ of the state space so that the unitaries $O, O_\star$ act on a total of 4 registers of size $m=\log|\mathbb S|$.}
    \label{fig:W}
\end{figure*}

\thirdreview{Given quantum circuits for $O$ and $O_\star$, we are now able to construct the Szegedy walk operators of $\mathcal P$ and $\mathcal P^\star$. However, Theorem~\ref{the:qubitized_eigvals} is stated for an SPUE of the discriminant of a reversible Markov kernel, whereas the dual kernel $\mathcal P=\mathcal{TA}$ is not reversible in general. Hermitianization combines the encodings of $\mathcal P$ and $\mathcal P^\star$ into a single SPUE of the discriminant of the reversible dilation
\begin{equation}
\overline{\mathcal P}=\ket{0}\bra1\otimes \mathcal P+\ket1\bra0\otimes\mathcal P^\star.
\end{equation}
Proposition~\ref{prop:overline_dual_P} provides such a construction (see~\cite{claudon2025quantumspeedupnonreversiblemarkov} for more details on quantum algorithms for nonreversible Markov chains). Figure~\ref{fig:W} illustrates the corresponding qubitized walk operator.}

\begin{Prop}
The Hermitianization $(\mathcal U, \boxtimes)$ of $(S, \square_\star, \square)$ is a SPUE of the discriminant $\overline{\mathcal D}$ of the reversible dilation $\overline{\mathcal P}=\ket{0}\bra1\otimes \mathcal P+\ket1\bra0\otimes\mathcal P^\star$ of $\mathcal P$. $\overline{\mathcal P}$ has a single stationary distribution $\frac12\begin{pmatrix}\pi & \pi\end{pmatrix}$ and a spectral gap of $1-\sqrt{1-\delta_\star}$, where $\delta_\star$ is the spectral gap of $\mathcal P\mathcal P^\star$.
\label{prop:overline_dual_P}
\end{Prop}

\begin{proof}
Using Propositions~\ref{prop:construct_dual} and~\ref{prop:construct_dual_star}, we can implement the Szegedy walk operators $2\square\square^\dagger-1$ of $\mathcal P$ and $2\square_\star\square_\star^\dagger-1$ of $\mathcal P^\star$. Notice that the Hermitianization 
\begin{equation}
(\mathcal U, \boxtimes) = \left(X\otimes S, \ket0\bra0\otimes \square_\star+\ket1\bra1\otimes \square\right)
\end{equation}
(see the definition in Proposition~\ref{prop:hermitianization}) of $(S, \square_\star, \square)$ is a SPUE of:
\begin{equation}
\overline{\mathcal D}=\begin{pmatrix}
0 & \mathcal D\\
\mathcal D^\dagger & 0
\end{pmatrix},    
\end{equation}
where $\mathcal D=\square_\star^\dagger S\square$ is the discriminant of $\mathcal P$ (see Definition~\ref{def:discriminant}). Also note that $\mathcal D^\dagger=\square^\dagger S\square_\star$ is the discriminant of $\mathcal P^\star$. Therefore, $\overline{\mathcal D}$ is the discriminant of
\begin{equation}
\overline{\mathcal P}=\begin{pmatrix}
0 & \mathcal P\\
\mathcal P^\star & 0
\end{pmatrix}.
\end{equation}
Proposition~\ref{prop:hermitianization} ensures that the eigenvalues of $\overline{\mathcal D}$ are plus or minus the square root of the eigenvalues of $\mathcal P\mathcal P^\star$ (i.e. the singular values of $\mathcal P$ with respect to $\langle\cdot, \cdot\rangle_\nu$). Moreover, $\ket+\ket\nu$ is clearly an eigenvector with eigenvalue $1$. 
\end{proof}

We want to conclude, using Theorem~\ref{the:qubitized_eigvals}, that the qubitized walk operator $\mathcal W$ of $(\mathcal U, \boxtimes)$ has angular gap $\Omega\left(\sqrt\delta\right)$. 

\subsection{Gap of the qubitized walk operator}\label{sec:gap}

The purpose of this subsection is to show that the qubitized walk operator has a quadratically amplified spectral gap. We prove this result by showing that $\delta_\star\in \Omega(\delta)$. Corollary~\ref{cor:gap_glauber} states the result in the Glauber choice case. The general case requires considering the lazy version of $P$ and is given by Corollary~\ref{cor:eiggap_vs_MHgap_metropolis}.

\subsubsection{The Glauber case}\label{sec:glauber}

\begin{Cor}
Under the Glauber choice, the spectral gap $\delta_\star$ of $\mathcal P\mathcal P^\star$ is equal to that of $P$: $\delta=\delta_\star$.
\label{cor:gap_glauber}
\end{Cor}

\begin{proof}
Under the Glauber choice, $\mathcal A=\mathcal A^2$. This follows from the fact that $A(x, y)+A(y, x)=1$, which implies: 
\begin{equation}
\begin{split}
&\begin{pmatrix}
\mathcal A((x,y), (x, y))&\mathcal A((x,y), (y, x))\\
\mathcal A((y,x), (x, y))&\mathcal A((y, x), (y, x))
\end{pmatrix}    
\\
&=
\begin{pmatrix}
1-A(x, y) & A(x, y)\\
A(y, x) & 1-A(y, x)
\end{pmatrix}\\
&=\begin{pmatrix}
1-A(x, y) & A(x, y)\\
1-A(x, y) & A(x, y)
\end{pmatrix}\\
&=\begin{pmatrix}
1-A(x, y) & A(x, y)\\
1-A(x, y) & A(x, y)
\end{pmatrix}^2,
\end{split}
\end{equation}
for every $(x,y)\in \mathcal S$. $\mathcal A$ is a direct sum of perfectly mixed Markov kernels. Therefore, $\mathcal P\mathcal P^\star=\mathcal{T}\mathcal A^2\mathcal T=\mathcal{TAT}$. Conclude with Corollary~\ref{cor:equal_mixing}.
\end{proof}

We now obtain an upper bound for the second largest eigenvalue of $\overline{\mathcal P}$.

\begin{Cor}
The second largest eigenvalue $\lambda$ of $\overline{\mathcal P}$ satisfies $\lambda=\sqrt{1-\delta_\star}$. In particular, 
\begin{equation}
1-\lambda= 1-\sqrt{1-\delta_\star}\geq\frac{\delta}2-\mathcal O\left(\delta^2\right),
\label{eq:eiggap_vs_MHgap}
\end{equation}
where $\delta$ is the spectral gap of the Metropolis-Hastings kernel $P$ with Glauber choice.
\label{cor:eiggap_vs_MHgap}
\end{Cor}

\begin{proof}
The second largest eigenvalue of $\overline{\mathcal P}$ is the second largest singular value of $\mathcal P$ (with respect to the scalar product $\braket{\cdot, \cdot}_\nu$). By definition, it is the square root of the second largest eigenvalue of $\mathcal P\mathcal P^\star$. Thus, $\lambda =\sqrt{1-\delta_\star}$ and the equality of Equation~\ref{eq:eiggap_vs_MHgap} holds true. The inequality follows from Corollary~\ref{cor:gap_glauber}.
\end{proof}

\subsubsection{General acceptance matrix}

Under the Metropolis choice and when targeting the uniform distribution with a symmetric proposal kernel, $\mathcal A^2=1$. As a consequence, 
\begin{equation}
\mathcal P\mathcal P^\star=\mathcal T^2=\mathcal T.
\end{equation}
In particular, $\mathcal P\mathcal P^\star$ is not ergodic. By replacing $A$ by $A/2$, namely implementing the lazy Metropolis-Hastings kernel, we can ensure that $\mathcal P\mathcal P^\star$ is ergodic with a mixing time equal to that of $P$, up to a multiplicative constant. In this subsection, $\mathcal P$ will always denote the dual kernel of $(1+P)/2$.

\begin{Prop}
Let $\delta$ be the spectral gap of a reversible ergodic Metropolis-Hastings kernel $P$. Let $\mathcal P$ be the dual kernel of $(1+P)/2$. Then, the spectral gap $\delta_\star$ of $\mathcal P\mathcal P^\star$ satisfies:
\begin{equation}
\delta_\star\geq \frac{\delta}{2}.
\end{equation}
\label{prop:deltastargeneral}
\end{Prop}

\begin{proof}
Since $\mathcal P=\mathcal T\left(\frac{1+\mathcal A}2\right)$, 
\begin{equation}
\begin{split}
\mathcal P\mathcal P^\star&=\mathcal T\left(\frac{1+\mathcal A}2\right)^2\mathcal T\\
&=\frac12\mathcal T\mathcal A\mathcal T+\frac12\mathcal T\left(\frac{1+\mathcal A^2}{2}\right)\mathcal T.
\end{split}
\end{equation}
Recall from Corollary~\ref{cor:equal_mixing} that $\mathcal{TAT}$ has a spectral gap $\delta$. Let $f:\mathcal S\to\mathbb C$ be such that $\langle f, f\rangle_\nu=1$ and $\nu f=0$. The variational principle gives:
\begin{equation}
\left\langle f, \mathcal T\mathcal A\mathcal T f\right\rangle_\nu\leq 1-\delta.
\end{equation}
$\mathcal T\left(\frac{1+\mathcal A^2}{2}\right)\mathcal T$ is also reversible and such that:
\begin{equation}
\left\langle f, \mathcal T\left(\frac{1+\mathcal A^2}{2}\right)\mathcal Tf\right\rangle_{\nu}\leq 1.
\end{equation}
Then, 
\begin{equation}
\begin{split}
\left\langle f, \mathcal P\mathcal P^\star f\right\rangle_\nu&=\frac12\left\langle f, \mathcal T\mathcal A\mathcal T f\right\rangle_\nu
+\frac12\left\langle f, \mathcal T\left(\frac{1+\mathcal A^2}{2}\right)\mathcal Tf\right\rangle_{\nu}\\
&\leq \frac12\left(1-\delta\right)+\frac12\\
&=1-\frac{\delta}{2}.
\end{split}
\end{equation}
Maximizing over all possible $f$, the variational principle implies that the spectral gap $\delta_\star$ of $\mathcal P\mathcal P^\star$ satisfies:
\begin{equation}
\delta_\star\geq \frac{\delta}{2}.
\end{equation}
\end{proof}

We now obtain an upper bound for the second largest eigenvalue of $\overline{\mathcal P}$.

\begin{Cor}
The second largest eigenvalue $\lambda$ of $\overline{\mathcal P}$ satisfies $\lambda=\sqrt{1-\delta_\star}$. In particular, 
\begin{equation}
1-\lambda= 1-\sqrt{1-\delta_\star}\geq\frac{\delta}4-\mathcal O\left(\delta^2\right),
\label{eq:eiggap_vs_MHgap2}
\end{equation}
where $\delta$ is the spectral gap of the Metropolis-Hastings kernel $P$.
\label{cor:eiggap_vs_MHgap_metropolis}
\end{Cor}

\begin{proof} Repeat the steps of the proof of Corollary~\ref{cor:eiggap_vs_MHgap}, and use Proposition~\ref{prop:deltastargeneral} for the inequality.
\end{proof}

\subsubsection{Main result}

We have found \thirdreview{an upper} bound for the second largest eigenvalue of $\overline{\mathcal P}$ for a general MH kernel. We will now turn this result into a lower bound for the angular gap of the qubitized walk operator $\mathcal W$.

\begin{The}
Let $P$ be an ergodic reversible Metropolis-Hastings kernel with proposal kernel $T$, acceptance matrix $A$, stationary distribution $\pi$ and spectral gap $\delta$. Using a constant number of uses of $O_T$ and $O_A$, we can construct a unitary $\mathcal W$ with unique $1$-eigenvector
\begin{equation}
\boxtimes\ket{+}\ket\nu\ket0,
\end{equation}
in the range of a partial isometry $\boxtimes$, the reflection through which can also be implemented with \thirdreview{two calls to each of $O_T, O_T^\dagger, O_A$ and $O_A^\dagger$}. Moreover, $\mathcal W$ has an angular gap in $\Omega\left(\sqrt\delta\right)$. 
\label{the:summary}
\end{The}

\begin{proof}
\thirdreview{Let us define  
\begin{equation}
\left\{\begin{aligned}
&2\boxtimes \boxtimes^\dagger-1&=\ket0\bra0\otimes \left(2\square_\star\square_\star^\dagger-1\right)\\
&&+\ket1\bra1\otimes\left(2\square\square^\dagger-1\right),\\
&\mathcal U=X\otimes S.&
\end{aligned}
\right.
\end{equation}
By definition, $(\mathcal U, \boxtimes)$ is the Hermitianization of the PUE $(S, \square_\star,\square)$, with specific properties detailed by Proposition~\ref{prop:hermitianization}.} Propositions~\ref{prop:construct_dual},~\ref{prop:construct_dual_star} and Equation~\ref{eq:reflection_quantumstep} show how to construct $2\square_\star\square_\star^\dagger-1$ and $2\square\square^\dagger-1$ with a constant number of uses of $O_T$ and $O_A$. $2\boxtimes\boxtimes^\dagger-1$ is a product of their $\ket0$ and $\ket1$ controlled versions. $\mathcal U$ is the tensor product of an $X$ gate on the ancilla and a register SWAP operator. The qubitized walk operator $\mathcal W=\left(2\boxtimes\boxtimes^\dagger-1\right)\mathcal U$ can therefore be implemented as shown in Figure~\ref{fig:W}, from a constant number of $O_A$ and $O_T$. In particular, a full application of $\mathcal{W}$ uses two calls to each of $O_A, O_A^\dagger, O_T, O_T^\dagger$. That its only $1$-eigenvector in the range of $\boxtimes$ is $\boxtimes\ket{+, \nu}$ is a consequence of Proposition~\ref{prop:overline_dual_P} and Theorem~\ref{the:qubitized_eigvals}. Corollary~\ref{cor:eiggap_vs_MHgap_metropolis} states that $\overline{\mathcal P}$ (and therefore $\overline{\mathcal D}$) has a spectral gap in $\Omega(\delta)$. Theorem~\ref{the:qubitized_eigvals} finally ensures that the angular gap of $\mathcal W$ is in $\Omega\left(\sqrt{\delta}\right)$. 
\end{proof}

Proposition~\ref{prop:pi_and_plusnu} shows that we can implement a unitary mapping $\boxtimes\ket{+, \nu}$ to $\ket\pi$ with a constant number of calls to $O_T$ and $O_A$.

\begin{Prop}
The following equation holds:
\begin{equation}
\begin{split}
&\ket+\ket\pi\ket{0,0}\\
&=O_T^\dagger O^\dagger\left(\ket0\bra0\otimes S+\ket1\bra1\otimes 1\right)\boxtimes \ket{+, \nu}.
\end{split}
\end{equation}
Here, $O_T^\dagger$ means $1\otimes O_T^\dagger\otimes 1$ and $O^\dagger$ means $1\otimes O^\dagger$.
\label{prop:pi_and_plusnu}
\end{Prop}

\begin{proof}
Since $\bra \nu\left(\square_\star^\dagger S\square\ket\nu\right)=\bra\nu(\mathcal D\ket\nu)=1$, we have $\square_\star\ket\nu=S\square\ket\nu$ and:
\begin{equation}
\begin{split}
\boxtimes \ket{+, \nu}&=\frac1{\sqrt2}\left(\ket0\otimes \square_\star\ket\nu+\ket1\otimes \square\ket\nu\right)\\
&=\frac1{\sqrt2}\left(\ket0\otimes S\square\ket\nu+\ket1\otimes \square\ket\nu\right)\\
&=\left(\ket0\bra0\otimes S+\ket1\bra1\otimes 1\right)\ket +\square\ket\nu.
\end{split}
\end{equation}
Thus, 
\begin{equation}
\left(\ket0\bra0\otimes S+\ket1\bra1\otimes 1\right)\boxtimes \ket{+, \nu}=\ket +\square\ket\nu.
\end{equation}
Because $(O, 1, \ket0)$ is a PUE of $\square$, we also have $\square\ket\nu=O\ket\nu\ket0$. In particular,
\begin{equation}
\ket+\ket\nu\ket0=O^\dagger\left(\ket0\bra0\otimes S+\ket1\bra1\otimes 1\right)\boxtimes \ket{+, \nu},
\end{equation}
(writing $O^\dagger$ for $1\otimes O^\dagger$) and we can prepare $\boxtimes\ket{+, \nu}$ from $\ket{+, \nu}$. The first marginal of $\nu$ is $\pi$, so preparing $\ket\nu$ is sufficient to sample from $\pi$. In order to prepare precisely $\ket\pi$, note that:
\begin{equation}
\ket\nu=\sum_{(x, y)\in \mathcal S}\sqrt{\pi(x)T(x, y)}\ket{x, y}=O_T\ket{\pi}\ket0.
\end{equation}
Therefore, writing $O_T^\dagger$ for $1\otimes O_T^\dagger\otimes 1$, 
\begin{equation}
\begin{split}
&\ket+\ket\pi\ket{0,0}\\
&=O_T^\dagger O^\dagger\left(\ket0\bra0\otimes S+\ket1\bra1\otimes 1\right)\boxtimes \ket{+, \nu}.
\end{split}
\end{equation}
\end{proof}

\begin{algorithm}[H]
\caption{Sampling from $\pi$}
\label{alg:sample_pi}

\KwIn{Oracles $O_T, O_A$ for a MH kernel $P$, and a state $\ket{\psi_0}\in\mathrm{Im}(\boxtimes)$ overlapping with $\boxtimes\ket{+}\ket{\nu}\ket{0}$.}
\KwOut{A sample $x\in\mathbb{S}$ distributed according to the stationary distribution $\pi$.}
\KwRes{For $m=\lceil\log_2|\mathbb S|\rceil$, each application of $\mathcal W$ acts on $4m+3$ qubits and uses two calls to each of $O_T,O_T^\dagger,O_A,O_A^\dagger$. The projection step uses $\widetilde O(1/\sqrt\delta)$ applications of $\mathcal W$, with additional phase-estimation ancillas.}
\BlankLine
Construct $O_{\mathcal A}$ from $O_A$, then $O,O_\star$ from $O_T$ and $O_{\mathcal A}$.\;
Implement $\mathcal W=(2\boxtimes\boxtimes^\dagger-1)(X\otimes S)$.\;
Project $\ket{\psi_0}$ onto the zero-phase eigenspace of $\mathcal W$, obtaining $\boxtimes\ket{+}\ket\nu\ket0$.\;
Apply $O_T^\dagger O^\dagger\left(\ket0\bra0\otimes S+\ket1\bra1\otimes 1\right)$ obtaining $\ket{+}\ket{\pi}\ket{0,0}$.\;
Measure the $\ket{\pi}$ register in the computational basis and return the outcome $x$.\;
\end{algorithm}

\thirdreview{Theorem~\ref{the:summary} and Proposition~\ref{prop:pi_and_plusnu} together complete the construction. Now, we can use these results to produce samples of $\pi$. Algorithm~\ref{alg:sample_pi} presents the procedure and its resource requirements.}
\thirdreview{Once we have access to $\mathcal{W}$, we implement the projection in Step~3 by standard quantum phase estimation or its improved variants~\cite{Kitaev1995QuantumMA, Ni2023lowdepthalgorithms}. This requires an initial state $\ket{\psi_0}\in\mathrm{Im}(\boxtimes)$ with non-zero overlap with the target state $\boxtimes\ket{+}\ket\nu\ket0$. The condition $\ket{\psi_0}\in\mathrm{Im}(\boxtimes)$ ensures that the state is in the relevant qubitization space, so phase estimation cannot project onto other potential zero-phase eigenvectors of $\mathcal W$. Within this subspace, nonzero phases are separated from the zero phase by the angular gap $\Delta=\Omega\left(\sqrt\delta\right)$ by Theorem~\ref{the:summary}. Resolving phases to precision $O(\Delta)$ requires $\widetilde O(1/\Delta)=\widetilde O\left(1/\sqrt\delta\right)$ calls to $\mathcal{W}$, which yields the quadratic quantum speedup. Measuring the phase register and postselecting the zero outcome prepares $\boxtimes\ket{+}\ket\nu\ket0$ up to the sampling overhead due to the initial-state overlap and the phase-estimation error. Proposition~\ref{prop:pi_and_plusnu} then maps the state to $\ket{+}\ket\pi\ket{0,0}$ with a constant number of calls to $O_T$ and $O_A$. Finally, measuring the state register in the computational basis returns a sample from $\pi$.}
\thirdreview{Beyond the asymptotic scaling, the reduced ancilla count and constant oracular cost make classical simulation  of small MH sampling instances possible. Section~\ref{sec:use_case} demonstrates this on a concrete MALA example, verifying the quadratic gap and showing that the proposed method can be analyzed numerically at small scale.}

\section{Conclusion}



\secondreview{In this work, we have presented new quantum circuits for the Metropolis-Hastings algorithm. We did so by implementing the quantum step operator of a dual Markov process, which lives on directed edges of the state space, avoiding the evaluation of rejection probabilities by classical arithmetic on qubit registers. 
Similarly, we constructed the quantum step operator of the time reversal of the dual process. These step operators were used to construct the Szegedy walk associated with the dual process. We have shown that this walk presents quadratic gap amplification and that its unique $1$-eigenvector encodes the MH stationary distribution. The step operators presented are of independent interest: for instance, they can be used to perform quantum singular value transformations. Moreover, the idea of using a process on the edges of the state space may be useful in other quantum circuit optimization problems. 
Additionally, we have presented an alternative approach to Szegedy's (Appendix~\ref{app:other}) that achieves the same quadratic gap amplification with even fewer registers. Our results show that small instances of the Metropolis-Hastings algorithm can be implemented on current quantum computers. It is an incremental step towards a full quadratic quantum speedup of a Markov Chain Monte-Carlo algorithm.}



\begin{appendix}

\section{\secondreview{Controlled-SWAP Symmetric Projected Unitary Encoding}}\label{app:other}

The authors of \cite{lemieux2020efficient} modified Szegedy's quantization procedure by replacing the register SWAP unitary by a controlled application of the proposed move. Such replacement removes the need for the quantum step operator of the Metropolis-Hastings walk. However, it is precisely this operator which is non-trivial to implement without additional qubits or gates different from $O_A$ and $O_T$. Following their approach, we obtain a $2m+1$-qubit-construction of a SPUE of the MH discriminant. The construction relies on replacing the SWAP operator by a controlled-SWAP, and is described in the following proposition.

\begin{Prop}
Let $P$ be a Metropolis-Hastings Markov kernel with proposal quantum step operator $O_T$ and acceptance operator $O_A$. Then,
$(O_T^\dagger O_A^\dagger S^c O_A O_T, \ket{0,0})$ is a SPUE of its discriminant $\mathcal D$:
\begin{equation}
\braket{0, 0|O_T^\dagger O_A^\dagger S^c O_A O_T|0, 0}=\mathcal D.
\end{equation}
Here, $S^c$ denotes a SWAP operator controlled by the coin qubit of $O_A$.
\label{prop:2m-construction}
\end{Prop}

\begin{proof}
Let $x\in \mathbb S$. Then, 
\begin{equation}
\begin{split}
&O_AO_T\ket{x, 0, 0} = O_A\sum_{y\in \mathbb S}\sqrt{T(x, y)}\ket{x, y, 0}\\
&=\sum_{y\in \mathbb S}\sqrt{T(x, y)A(x, y)}\ket{x, y, 1}\\
&+\sum_{y\in \mathbb S}\sqrt{T(x, y)(1-A(x, y))}\ket{x, y, 0}.
\end{split}
\label{eq:forward}
\end{equation}
Applying $S^c$, we obtain:
\begin{equation}
\begin{split}
&S^cO_AO_T\ket{x, 0, 0} =\sum_{y\in \mathbb S}\sqrt{T(x, y)A(x, y)}\ket{y, x, 1}\\
&+\sum_{y\in \mathbb S}\sqrt{T(x, y)(1-A(x, y))}\ket{x, y, 0}.
\end{split}
\label{eq:forward_and_cS}
\end{equation}
Let $z\in \mathbb S$. Combining Equations~\ref{eq:forward} and~\ref{eq:forward_and_cS}, 
\begin{equation}
\begin{split}
&\braket{x, 0, 0|O_T^\dagger O_A^\dagger S^c O_A O_T|z, 0, 0}\\
&=\sum_{y, t\in \mathbb S}\sqrt{T(x, y)A(x, y)T(z, t)A(z, t)}\braket{y, x|z, t}\\
&+\sum_{y, t\in \mathbb S}\sqrt{T(x, y)(1-A(x, y))T(z, t)(1-A(z, t))}\braket{x, y|z, t}\\
&=\sqrt{T(x, z)A(x, z)T(z, x)A(z, x)}\\
&+\sum_{t\in \mathbb S}T(z, t)(1-A(z, t))\braket{x| z}\\
&=\sqrt{P(x, z)P(z, x)}\\
&=\mathcal D(x, z).
\end{split}
\end{equation}
\end{proof}

Still, our approach using a process on the edges provides an alternative solution that remains closer to Szegedy's construction. Because it builds the quantum step operator of a Markov chain on the edges with transitions of the tails following exactly the MH algorithm, it provides a tool of independent interest.

\section{Metropolis-Adjusted Langevin Algorithm: the circuits}\label{app:OAOT}

In this Appendix, we present constructions for the proposal $O_T$ and acceptance $O_A$ operators associated with the discretized MALA algorithm presented in Section~\ref{sec:use_case}. We emphasize that the circuit does not require using qubit registers as classical bits on which we perform arithmetic operations. We also count the number of required ancilla qubits. For the remainder of this section, the discrete state space is $\mathbb S=\{k/2^n\}_{k=0}^{2^n-1}$, where $n$ is a positive integer, and we assume periodic boundary conditions. We will make use of single-ancilla efficient implementations of diagonal operators of the form
\begin{equation}
\sum_{x\in \mathbb S}e^{2\pi i f(x)}\ket x\bra x,   
\end{equation}
and
\begin{equation}
\sum_{x, y\in \mathbb S}e^{2\pi i f(x, y)}\ket{x, y}\bra{x, y},
\end{equation}
for smooth functions $f$, as presented in \cite{Welch_2014, zylberman_diag}.

\subsubsection{Gaussian state preparation}

There are numerous methods to implement Gaussian states on quantum computers \cite{kuklinski2025simplergaussianstatepreparation, xie2025efficientgaussianstatepreparation, Rattew2021efficient, zylberman_stateprep}. We quote the result from \cite{zylberman_stateprep}, which is based on the efficient implementation of diagonal operators. For $f:[0, 1]\to[0,1]$, denote by $\ket f$ the state:
\begin{equation}
\ket f=\sum_{x\in \mathbb S}f(x)\ket x.
\end{equation}

\begin{Prop}[\cite{zylberman_stateprep}]
Let $f:[0, 1]\to[0, 1]$ be a differentiable function such that $\braket{f|f}=1$, and $\epsilon \in ]0, \pi/\|f\|_\infty]$. There is a circuit $U$ of size $\mathcal O(n+1/\sqrt\epsilon)$, using one ancilla qubit, such that
\begin{equation}
\left\|\frac{U\ket 0}{\|U\ket0\|}-\ket f\right\|\leq \epsilon,
\end{equation}
and such that
\begin{equation}
\|U\ket0\|^2=\Theta(\epsilon).
\end{equation}
\label{prop:zsp}
\end{Prop}

Using amplitude amplification with an additional qubit, Proposition~\ref{prop:zsp} leads to a $2$-ancilla efficient Gaussian state preparation.

\subsection{Gradient descent operator}

The MALA algorithm suggests a new move according to a Gaussian centered on the gradient descent proposal. We would like to make use of an operator performing:
\begin{equation}
\ket{x, y}\to\ket{x, y+x-\tau\beta\nabla U(x)}, 
\end{equation}
for every $x, y\in \mathbb S$. When $\nabla U$ is smooth, we can efficiently implement
\begin{equation}
U_\nabla=\sum_{x\in \mathbb S}e^{-2\pi i(x-\tau\beta\nabla U(x))}\ket x\bra x.
\label{eq:Unabla}
\end{equation}

Combined with the Quantum Phase Estimation algorithm (Proposition~\ref{prop:qpe}), it leads to the desired operator.

\begin{Prop}[Quantum Phase Estimation \cite{Kitaev1995QuantumMA, Nielsen_Chuang_2010}]
Let $f:\mathbb S\to\mathbb S$ and $U_f$ be the unitary such that, for every $x, y\in \mathbb S$:
\begin{equation}
U_f\ket{x, y}=\ket{x, y+f(x)}.
\end{equation}
Then, $U_f$ is the Quantum Phase Estimation circuit of $V_f$ defined as:
\begin{equation}
V_f=\sum_{x\in \mathbb S}e^{2\pi i f(x)}\ket x\bra x.
\end{equation}
\label{prop:qpe}
\end{Prop}

\subsection{Proposal operator $O_T$}

The Langevin algorithm proposal operator satisfies
\begin{equation}
O_T\ket{x, 0}\propto\sum_{z\in \mathbb S}e^{-\frac{z^2}{8\tau}}\ket{x, x-\tau\beta\nabla U(x)+z},
\end{equation}
for every $x\in \mathbb S$. It can be implemented by first preparing a Gaussian state with the right standard deviation on the second state register, and then applying the gradient descent shift $U_\nabla$ such that
\begin{equation}
U_\nabla \ket{x, y}=\ket{x, y+x-\tau\beta\nabla U(x)},
\end{equation}
for every $x, y\in \mathbb S$.

\begin{Prop} Let $O_g$ denote the Gaussian state preparation circuit with 2 ancilla qubits. Then,
\begin{equation}
O_T = U_\nabla O_g
\end{equation}
is a valid proposal operator.
\end{Prop}

\begin{proof} Let $x\in \mathbb S$ and compute:
\begin{equation}
\begin{split}
&U_\nabla O_g\ket{x, 0} \propto U_\nabla\sum_{z\in \mathbb S}e^{-\frac{z^2}{8\tau}}\ket{x, z}\\
&=\sum_{z\in \mathbb S}e^{-\frac{z^2}{8\tau}}\ket{x, x-\tau\beta\nabla U(x)+z}.
\end{split}
\end{equation}
\end{proof}

Overall, $O_T$ requires no more than the two ancillae that are required to construct $O_g$.

\subsection{Acceptance operator $O_A$}

In order to implement $O_A$, let us first diagonalize it.

\begin{Prop}[Diagonalization]
$O_A$ can be diagonalized with $4$ single-qubit gates.
\end{Prop}

\begin{proof}
By definition of $O_A$,
\begin{equation}
O_A=\bigoplus_{x, y\in \mathbb S}\ket{x, y}\bra{x, y}\otimes \exp(i\theta(x, y)Y),
\end{equation}
with $\theta(x, y)=-\sin^{-1}\left(\sqrt{A(x, y)}\right)$. Recalling that $Y=SHZHS^\dagger$, where $S=\ket0\bra0+i\ket1\bra1$ and $H$ is the Hadamard gate,
\begin{equation}
O_A = SH\exp\left(i\sum_{x, y\in \mathbb S}\theta(x, y)\ket{x, y}\bra{x, y}\otimes Z\right)HS^\dagger,
\end{equation}
where $S, S^\dagger, H$ act on $O_A$'s coin qubit.
\end{proof}

Whenever $\theta(x, y)=-\sin^{-1}\left(\sqrt{A(x, y)}\right)$ is approximated by a smooth function (eventually restricted to $(x, y)\in \mathcal S$), the techniques presented in~\cite{Welch_2014, zylberman_diag} lead to a single-ancilla efficient implementation of $O_A$.

\end{appendix}

\section*{Acknowledgements}

This work has been funded by the European Research Council (ERC) under the European Union’s Horizon 2020 research and innovation program (grant No 810367), project EMC2 (J.-P. P.). Support from the PEPR EPIQ - Quantum Software (ANR-22-PETQ-0007, J.-P. P.) and HQI (J.-P. P.) programs is acknowledged. P.R.-R. received the support of a fellowship from "la Caixa" Foundation (ID 100010434), with fellowship code LCF/BQ/EU23/12010085.
\onecolumngrid\newpage
\section*{References}
\bibliography{references}

@misc{chiang2009efficientcircuitsquantumwalks,
      title={Efficient Circuits for Quantum Walks}, 
      author={Chen-Fu Chiang and Daniel Nagaj and Pawel Wocjan},
      year={2009},
      eprint={0903.3465},
      archivePrefix={arXiv},
      primaryClass={quant-ph},
      url={https://arxiv.org/abs/0903.3465}, 
}

@article{Ni2023lowdepthalgorithms,
  doi = {10.22331/q-2023-11-06-1165},
  url = {https://doi.org/10.22331/q-2023-11-06-1165},
  title = {On low-depth algorithms for quantum phase estimation},
  author = {Ni, Hongkang and Li, Haoya and Ying, Lexing},
  journal = {{Quantum}},
  issn = {2521-327X},
  publisher = {{Verein zur F{\"{o}}rderung des Open Access Publizierens in den Quantenwissenschaften}},
  volume = {7},
  pages = {1165},
  month = nov,
  year = {2023}
}

@book{Nielsen_Chuang_2010, place={Cambridge}, title={Quantum Computation and Quantum Information: 10th Anniversary Edition}, publisher={Cambridge University Press}, author={Nielsen, Michael A. and Chuang, Isaac L.}, year={2010}, nolink={}}

@article{Loke_2017,
   title={Efficient quantum circuits for Szegedy quantum walks},
   volume={382},
   ISSN={0003-4916},
   url={http://dx.doi.org/10.1016/j.aop.2017.04.006},
   DOI={10.1016/j.aop.2017.04.006},
   journal={Annals of Physics},
   publisher={Elsevier BV},
   author={Loke, T. and Wang, J.B.},
   year={2017},
   month=jul, pages={64–84} }

@article{PhysRevA.86.042338,
  title = {Efficient circuit implementation of quantum walks on non-degree-regular graphs},
  author = {Loke, T. and Wang, J. B.},
  journal = {Phys. Rev. A},
  volume = {86},
  issue = {4},
  pages = {042338},
  numpages = {7},
  year = {2012},
  month = {Oct},
  publisher = {American Physical Society},
  doi = {10.1103/PhysRevA.86.042338},
  url = {https://link.aps.org/doi/10.1103/PhysRevA.86.042338}
}

@misc{claudon2025quantumspeedupnonreversiblemarkov,
      title={Quantum Speedup for Nonreversible Markov Chains}, 
      author={Baptiste Claudon and Jean-Philip Piquemal and Pierre Monmarché},
      year={2025},
      eprint={2501.05868},
      archivePrefix={arXiv},
      primaryClass={quant-ph},
      url={https://arxiv.org/abs/2501.05868}, 
}

@article{https://doi.org/10.1155/2015/183918,
author = {Paquet, Eric and Viktor, Herna L.},
title = {Molecular Dynamics, Monte Carlo Simulations, and Langevin Dynamics: A Computational Review},
journal = {BioMed Research International},
volume = {2015},
number = {1},
pages = {183918},
doi = {https://doi.org/10.1155/2015/183918},
url = {https://onlinelibrary.wiley.com/doi/abs/10.1155/2015/183918},
eprint = {https://onlinelibrary.wiley.com/doi/pdf/10.1155/2015/183918},
year = {2015}
}

@misc{xie2025efficientgaussianstatepreparation,
      title={Efficient Gaussian State Preparation in Quantum Circuits}, 
      author={Yichen Xie and Nadav Ben-Ami},
      year={2025},
      eprint={2507.20317},
      archivePrefix={arXiv},
      primaryClass={quant-ph},
      url={https://arxiv.org/abs/2507.20317}, 
}

@article{Rattew2021efficient,
  doi = {10.22331/q-2021-12-23-609},
  url = {https://doi.org/10.22331/q-2021-12-23-609},
  title = {The {E}fficient {P}reparation of {N}ormal {D}istributions in {Q}uantum {R}egisters},
  author = {Rattew, Arthur G. and Sun, Yue and Minssen, Pierre and Pistoia, Marco},
  journal = {{Quantum}},
  issn = {2521-327X},
  publisher = {{Verein zur F{\"{o}}rderung des Open Access Publizierens in den Quantenwissenschaften}},
  volume = {5},
  pages = {609},
  month = dec,
  year = {2021}
}

@inproceedings{Ozols_2012, series={ITCS ’12},
   title={Quantum rejection sampling},
   url={http://dx.doi.org/10.1145/2090236.2090261},
   DOI={10.1145/2090236.2090261},
   booktitle={Proceedings of the 3rd Innovations in Theoretical Computer Science Conference},
   publisher={ACM},
   author={Ozols, Maris and Roetteler, Martin and Roland, Jérémie},
   year={2012},
   month=Jan, pages={290–308},
   collection={ITCS ’12} }

@misc{kuklinski2025simplergaussianstatepreparation,
      title={A simpler Gaussian state-preparation}, 
      author={Parker Kuklinski and Benjamin Rempfer and Kevin Obenland and Justin Elenewski},
      year={2025},
      eprint={2508.03987},
      archivePrefix={arXiv},
      primaryClass={quant-ph},
      url={https://arxiv.org/abs/2508.03987}, 
}

@article{Welch_2014,
   title={Efficient quantum circuits for diagonal unitaries without ancillas},
   volume={16},
   ISSN={1367-2630},
   url={http://dx.doi.org/10.1088/1367-2630/16/3/033040},
   DOI={10.1088/1367-2630/16/3/033040},
   number={3},
   journal={New Journal of Physics},
   publisher={IOP Publishing},
   author={Welch, Jonathan and Greenbaum, Daniel and Mostame, Sarah and Aspuru-Guzik, Alan},
   year={2014},
   month=mar, pages={033040} }

@article{zylberman_stateprep,
  title = {Efficient quantum state preparation with Walsh series},
  author = {Zylberman, Julien and Debbasch, Fabrice},
  journal = {Phys. Rev. A},
  volume = {109},
  issue = {4},
  pages = {042401},
  numpages = {22},
  year = {2024},
  month = {Apr},
  publisher = {American Physical Society},
  doi = {10.1103/PhysRevA.109.042401},
  url = {https://link.aps.org/doi/10.1103/PhysRevA.109.042401}
}

@article{zylberman_diag,
author = {Zylberman, Julien and Nzongani, Ugo and Simonetto, Andrea and Debbasch, Fabrice},
title = {Efficient Quantum Circuits for Non-Unitary and Unitary Diagonal Operators with Space-Time-Accuracy Trade-Offs},
year = {2025},
issue_date = {June 2025},
publisher = {Association for Computing Machinery},
address = {New York, NY, USA},
volume = {6},
number = {2},
url = {https://doi.org/10.1145/3718348},
doi = {10.1145/3718348},
journal = {ACM Transactions on Quantum Computing},
month = apr,
articleno = {15},
numpages = {43}
}

@article{10.1093/biomet/57.1.97,
    author = {Hastings, W. K.},
    title = {Monte Carlo sampling methods using Markov chains and their applications},
    journal = {Biometrika},
    volume = {57},
    number = {1},
    pages = {97-109},
    year = {1970},
    month = {04},
    issn = {0006-3444},
    doi = {10.1093/biomet/57.1.97},
    url = {https://doi.org/10.1093/biomet/57.1.97},
    eprint = {https://academic.oup.com/biomet/article-pdf/57/1/97/23940249/57-1-97.pdf},
}

@article{Rodenas_Ruiz_Claudon, title={qc-metropolis}, url={https://github.com/qrodenas/qc-metropolis}, journal={qc-metropolis}, author={Rodenas Ruiz, Pablo and Claudon, Baptiste}, year={2025}}

@article{doi:10.1126/science.220.4598.671,
author = {S. Kirkpatrick  and C. D. Gelatt  and M. P. Vecchi },
title = {Optimization by Simulated Annealing},
journal = {Science},
volume = {220},
number = {4598},
pages = {671-680},
year = {1983},
doi = {10.1126/science.220.4598.671},
URL = {https://www.science.org/doi/abs/10.1126/science.220.4598.671},
eprint = {https://www.science.org/doi/pdf/10.1126/science.220.4598.671}
}

@INPROCEEDINGS{1366222,
  author={Szegedy, M.},
  booktitle={45th Annual IEEE Symposium on Foundations of Computer Science}, 
  title={Quantum speed-up of Markov chain based algorithms}, 
  year={2004},
  volume={},
  number={},
  pages={32-41},
  doi={10.1109/FOCS.2004.53}}

@book{10.5555/1051461,
author = {Landau, David and Binder, Kurt},
title = {A Guide to Monte Carlo Simulations in Statistical Physics},
year = {2005},
isbn = {0521842387},
publisher = {Cambridge University Press},
address = {USA},
nolink = {}
}

@article{LOVASZ2006392,
title = {Simulated annealing in convex bodies and an O*(n4) volume algorithm},
journal = {Journal of Computer and System Sciences},
volume = {72},
number = {2},
pages = {392-417},
year = {2006},
note = {JCSS FOCS 2003 Special Issue},
issn = {0022-0000},
doi = {https://doi.org/10.1016/j.jcss.2005.08.004},
url = {https://www.sciencedirect.com/science/article/pii/S0022000005000966},
author = {László Lovász and Santosh Vempala}
}

@misc{sunderhauf2023generalizedquantumsingularvalue,
      title={Generalized Quantum Singular Value Transformation}, 
      author={Christoph Sünderhauf},
      year={2023},
      eprint={2312.00723},
      archivePrefix={arXiv},
      primaryClass={quant-ph},
      url={https://arxiv.org/abs/2312.00723}, 
}

@inproceedings{
childs2022quantum,
title={Quantum Algorithms for Sampling Log-Concave Distributions and Estimating Normalizing Constants},
author={Andrew Childs and Tongyang Li and Jin-Peng Liu and Chunhao Wang and Ruizhe Zhang},
booktitle={Advances in Neural Information Processing Systems},
editor={Alice H. Oh and Alekh Agarwal and Danielle Belgrave and Kyunghyun Cho},
year={2022},
url={https://openreview.net/forum?id=zofwPmKL-DO}
}

@misc{Brassard_2002,
   title={Quantum amplitude amplification and estimation},
   ISSN={0271-4132},
   url={http://dx.doi.org/10.1090/conm/305/05215},
   DOI={10.1090/conm/305/05215},
   journal={Quantum Computation and Information},
   publisher={American Mathematical Society},
   author={Brassard, Gilles and Høyer, Peter and Mosca, Michele and Tapp, Alain},
   year={2002},
   pages={53–74} }

@article{Yoder_2014,
   title={Fixed-Point Quantum Search with an Optimal Number of Queries},
   volume={113},
   ISSN={1079-7114},
   url={http://dx.doi.org/10.1103/PhysRevLett.113.210501},
   DOI={10.1103/physrevlett.113.210501},
   number={21},
   journal={Physical Review Letters},
   publisher={American Physical Society (APS)},
   author={Yoder, Theodore J. and Low, Guang Hao and Chuang, Isaac L.},
   year={2014},
   month=nov }

@article{Kitaev1995QuantumMA,
  title={Quantum measurements and the Abelian Stabilizer Problem},
  author={Alexei Y. Kitaev},
  journal={Electron. Colloquium Comput. Complex.},
  year={1995},
  volume={TR96},
  url={https://api.semanticscholar.org/CorpusID:17023060}
}

@misc{claudon2025simplealgorithmreflecteigenspaces,
      title={A simple algorithm to reflect through eigenspaces of unitaries}, 
      author={Baptiste Claudon},
      year={2025},
      eprint={2412.09320},
      archivePrefix={arXiv},
      primaryClass={quant-ph},
      url={https://arxiv.org/abs/2412.09320}, 
}

@article{lemieux2020efficient,
  doi = {10.22331/q-2020-06-29-287},
  url = {https://doi.org/10.22331/q-2020-06-29-287},
  title = {Efficient {Q}uantum {W}alk {C}ircuits for {M}etropolis-{H}astings {A}lgorithm},
  author = {Lemieux, Jessica and Heim, Bettina and Poulin, David and Svore, Krysta and Troyer, Matthias},
  journal = {{Quantum}},
  issn = {2521-327X},
  publisher = {{Verein zur F{\"{o}}rderung des Open Access Publizierens in den Quantenwissenschaften}},
  volume = {4},
  pages = {287},
  month = jun,
  year = {2020}
}

@book{levin2008markov,
  author = {Levin, D.A. and Peres, Y. and Wilmer, E.L.},
  isbn = {9780821886274},
  publisher = {American Mathematical Soc.},
  title = {Markov Chains and Mixing Times},
  url = {http://pages.uoregon.edu/dlevin/MARKOV/},
  year = 2008
}

\end{document}